\documentclass[superscriptaddress,twocolumn,amsmath,amssymb]{revtex4}
\usepackage{graphicx}
\usepackage{xcolor}
\usepackage{ulem}
\usepackage{braket}
\usepackage{mathtools}
\usepackage{enumitem}

\newcommand{\ketbra}[2]{\mathinner{|{#1}\rangle \! \langle{#2}|}}

\usepackage{hyperref}
\hypersetup{
colorlinks=true,
linkcolor=blue,
filecolor=magenta,
citecolor=red
}

\begin{document}

\title{Moment expansion method for composite open quantum systems \\
including a damped oscillator mode}

\author{Masaaki Tokieda}
\email{tokieda.masaaki.4e@kyoto-u.ac.jp}
\affiliation{Laboratoire de Physique de l'École Normale Supérieure, Inria, ENS, Mines Paris, PSL, CNRS, Université PSL, Paris, France}

\begin{abstract}
  We consider a damped oscillator mode that is resonantly driven and is coupled to an arbitrary target system via the position quadrature operator.
  For such a composite open quantum system, we develop a numerical method to compute the reduced density matrix of the target system and the low-order moments of the quadrature operators.
  In this method, we solve the evolution equations for quantities related to moments of the quadrature operators, rather than for the density matrix elements as in the conventional approach.
  The application to an optomechanical setting shows that the new method can compute the correlation functions accurately with a significant reduction in the computational cost.
  Since the method does not involve any approximation in its abstract formulation itself, we investigate the numerical accuracy closely.
  This study reveals the numerical sensitivity of the new approach in certain parameter regimes.
  We find that this issue can be alleviated by using the position basis instead of the commonly used Fock basis.
\end{abstract}

\maketitle


\section{Introduction}
\label{sec:Intro}

Open quantum systems have attracted attention from various perspectives.
One example is in the field of quantum technology, which has been rapidly growing in recent years.
In quantum technology applications, we typically consider a situation where a composite system consisting of a target system and a measurement or control apparatus interacts with the surrounding environment \cite{OpenQuantumSystems,QuantumNoise,QuantumMeasurement}.
When simulating such a composite open quantum system, one of the major difficulties is the dimensionality.
For an isolated quantum system, on the one hand, a state is represented by a vector in the Hilbert-space.
A state of an open quantum system, on the other hand, is represented by a density matrix, which is a square matrix with the Hilbert-space dimension.
Therefore, the number of degrees of freedom is linear with respect to the Hilbert-space dimension in simulations of an isolated quantum system, whereas it is quadratic for an open quantum system.
This results in a much larger, possibly prohibitive computational workload, especially for a composite system where the Hilbert-space dimension itself could be large.
Various methods have been proposed for efficient numerical simulations.
For a Lindblad equation, which is a Markovian master equation ensuring trace preservation and complete positivity of the time evolution map \cite{Lindblad}, methods that are applicable to general settings include the stochastic unraveling \cite{OpenQuantumSystems,Dalibard92,Gisin92} and the low rank approximation \cite{LowRank}.

In this article, we consider a composite open quantum system consisting of an arbitrary target system and an oscillator mode that serves as a measurement or control apparatus.
For later purposes, we introduce the trace operations over the total, target, and oscillator systems as ${\rm tr}$, ${\rm tr}_S$, and ${\rm tr}_{\rm osc}$, respectively.
When simulating such a system, the oscillator degrees of freedom in the total density matrix $\rho$ are commonly expanded using the Fock basis, which is the eigenbasis of the operator $a^\dagger a$ with $a$ and $a^\dagger$ the annihilation and creation operators of the oscillator mode, respectively.
Each matrix element is an operator on the target system space, and its temporal evolution can be determined by solving the corresponding master equation.
Using its solution, we can obtain the reduced density matrix of the target system, which is defined by ${\rm tr}_{\rm osc}(\rho)$.
The reduced density matrix allows us to evaluate any physical quantity associated with the target system.
We can also evaluate moments of the position $X_{\rm osc} = (a + a^\dagger)/\sqrt{2}$ and momentum $P_{\rm osc}= (a - a^\dagger) / (\sqrt{2} i)$ quadrature operators as $\{ \braket{X_{\rm osc}^m P_{\rm osc}^n} \}_{m,n=0,1,2,\dots}$ with $\braket{O} = {\rm tr} (O \rho)$ the expectation value of an operator $O$.
These quantities characterize the phase space distribution of the oscillator state.
For instance, the first-order moments ($\braket{X_{\rm osc}}$ and $\braket{P_{\rm osc}}$) determine the mean location in the phase space,
and the second-order moments ($\braket{X_{\rm osc}^2}$, $\braket{P_{\rm osc}^2}$, and $\braket{X_{\rm osc} P_{\rm osc}}$) together with the first-order moments determine the width of the distribution.

This article presents an alternative approach for simulating such a system.
In this approach, we track the evolution of quantities related to moments of the quadrature operators, rather than the density matrix elements as in the conventional approach.
The idea of using moments for composite open quantum systems comprising of an oscillator mode has already been discussed in the literature.
For instance, it has widely been applied to systems consisting solely of oscillator modes, where the generator is given by polynomials of the quadrature operators with a total degree not exceeding two.
In such systems, the evolution of moments up to the second-order is independent of the other higher-order moments \cite{Teretenkov19}.
Thus, one can easily obtain the evolution of the former, which are sufficient to fully characterize the oscillator state when the initial state is Gaussian.
Such cumulant expansion method can now be systematically implemented on arbitrary open quantum systems to any desired order using recently developed package QuantumCumulants.jl \cite{Plankensteiner22}.
The idea based on moments was also applied to an oscillator-atom system in \cite{Link22}.
The authors of that article considered quantities proportional to $\{ {\rm tr}_{\rm osc}(a^m \rho (a^\dagger)^n) \}_{m,n = 0,1,2,\dots}$.
Note that these are operators on the target system space, not $c$ numbers, and that taking their trace over the target system space yields the moments of the ladder operators.
The authors then discussed that the computations of the reduced density matrix ${\rm tr}_{\rm osc} (\rho)$, which corresponds to the element with $m = n = 0$, can be efficiently performed by tracking the evolution of those quantities.

Instead of ${\rm tr}_{\rm osc}(a^m \rho (a^\dagger)^n)$, we propose a new method based on a different representation, which provides moments of the quadrature operators.
While the new method (hereafter referred to as the moment expansion method) can be applied to general settings, it is especially advantageous as a numerical method when the oscillator mode is resonantly driven and is coupled to the target system via the position quadrature $X_{\rm osc}$.
Such systems have been discussed in optomechanical settings,
including quantum nondemolition measurement of the photon intensity \cite{Heidmann97}, back-action evading measurement of the mechanical oscillator displacement \cite{CFJ08}, observation of the quantized energy of the mechanical oscillator \cite{Hauer18} and generation of macroscopic-scale entanglement \cite{Mancini02,Clarke20}.
For such systems, the moment expansion method can facilitate the computation of the reduced density matrix and the low-order moments of the quadrature operators.
This is because the number of degrees of freedom in their computation scales linearly with respect to the truncated dimension of the oscillator state space, while it scales quadratically in the conventional method based on the density matrix.

Although the evolution of the total system (comprising the target and oscillator) is assumed to be governed by a Markovian Lindblad equation, the dynamics of the target system alone are generally non-Markovian \cite{Rivas14,Breuer16}.
For simulating non-Markovian dynamics, the bath oscillator model, which is a system-bath Hamiltonian model assuming the bath to be a collection of harmonic oscillators linearly coupled to the system, is widely used (see, e.g., \cite{Weiss08}).
Notably, there is a mathematical equivalence between a bath oscillator model and a Lindblad equation with fictitious oscillator modes, known as the pseudomode theory \cite{Imamoglu94,Garraway97,Tamascelli17}.
This equivalence allows us to view the Lindblad systems under investigation as a pseudomode description of non-Markovian dynamics.
In addition to the pseudomode theory, various methods exist for simulating the bath oscillator model, including non-Markovian quantum state diffusion \cite{Strunz96,Diosi98}, reaction coordinate mapping \cite{Cederbaum05,Chin10,Prior10}, path integral method \cite{Makri95_1,Makri95_2,Strathearn18}, and hierarchical equations of motion (HEOM) \cite{TK89,Yan04,Tanimura20}.
This article focuses on HEOM.
The HEOM simulations can be simplified when the bath correlation function is expressed as a sum of real exponential functions, as utilized in the original formulation \cite{TK89} and has since been recognized by numerous researchers, see e.g., \cite{Tang15,Suess15,Ikeda20}.
Building on the link established by the pseudomode theory described above, it is then expected that application of this simplification will facilitate simulation of Lindblad systems under similar conditions.
In this article, we see that the simplification is achievable by representing the total system state using quantities related to the moments of the quadrature operator.
The connection between HEOM and the pseudomode description was suggested in \cite{Lambert19} and recently established in \cite{Xu23}, where HEOM was shown to lead to a Lindblad-like equation after a specific similarity transformation.
Conversely, we derive HEOM starting from a Lindblad equation.
While the transformation introduced in \cite{Xu23} was abstract, we explore it based on quadrature operator moments and detail the simplification procedure from this perspective.

The article is organized as follows.
After presenting the formulation in Sec.$\,$\ref{sec:method}, we provide examples of its practical relevance upon applying it to the computation of the correlation function in an optomechanical system in Sec.$\,$\ref{sec:OptMech}.
We show that the moment expansion method provides the same quality of numerical results as the conventional method, at a largely reduced computational cost.
In addition to the efficiency compared with the conventional method, we also investigate the accuracy of the moment expansion method in Sec.$\,$\ref{sec:NumAccuracy}.
 While the intrinsic numerical sensitivity of the method is observed, we demonstrate that it can be mitigated by employing the eigenbasis of the position quadrature operator $X_{\rm osc}$.
Sec.$\,$\ref{sec:conc} contains some concluding remarks.


\section{Methodology}
\label{sec:method}

A composite system studied in this article consists of an arbitrary target system and an oscillator mode.
The total density matrix $\rho$ follows the master equation
\begin{equation}
  \frac{d}{dt} \rho = \mathcal{L} (\rho),
  \label{eq:method_Leq}
\end{equation}
where $\mathcal{L}$ is a Liouvillian given by
\begin{equation}
  \begin{gathered}
    \mathcal{L} (\rho) = \mathcal{L}_S (\rho) \\
    + [\epsilon a^\dagger - \epsilon^* a , \rho] + \kappa \mathcal{D}[a] (\rho) \\
    - i [ V_S ( a + a^\dagger), \rho].
  \end{gathered}
  \label{eq:method_Ltot}
\end{equation}
Throughout this article, we omit the tensor product symbol, $\otimes$ whenever it is evident and set $\hbar = 1$.
The superoperator $\mathcal{L}_S$ acts only on the target system operators nontrivially and describes internal dynamics of the target system.
Since the following formulation holds regardless of the detail of $\mathcal{L}_S$, we do not specify it until we discuss practical examples in later sections.
The second line describes the internal dynamics of the oscillator mode.
The commutator involving $\epsilon$ is a drive term with $\epsilon$ the drive amplitude.
The term with $\kappa$ describes a single-photon loss, where $\kappa$ is the loss rate and $\mathcal{D}[a](\rho) = a \rho a^\dagger - \{ a^\dagger a, \rho \}/2$ is the dissipator with the anticommutator $ \{ a^\dagger a, \rho \} = a^\dagger a \rho + \rho a^\dagger a$.
The coupling between the target system and the oscillator mode is described by the third line, with $V_S$ a Hermitian target system operator.

In Eq.$\,$(\ref{eq:method_Ltot}), we assume that the oscillator mode is resonantly driven.
For this reason, a commutator involving $a^\dagger a$, which represents the internal energy of the oscillator mode, is omitted.
We also make an assumption on the form of the interaction Hamiltonian, $V_S ( a + a^\dagger)$.
A general linear coupling form is given by $L_S \, a^\dagger + L_S^\dagger \, a$ with $L_S$ a target system operator that is not necessarily Hermitian.
In Eq.$\,$(\ref{eq:method_Ltot}), we assume that $L_S$ is equal to a Hermitian operator $V_S$.
Evolution equations for moments can be obtained without these assumptions, as detailed in Appendix \ref{app:RhoXiChi}.
In the main text, however, we concentrate on the master equation Eq.$\,$(\ref{eq:method_Ltot}) for which the new method presented in this article is particularly efficient.

Before discussing the detail of the moment expansion method, we review the conventional approach.
In this approach, the density matrix $\rho$ is expanded in the Fock basis.
Let $\ket{n} = ((a^\dagger)^n / \sqrt{n!}) \ket{0}$, for $n = 0,1,2,\dots$, denote the Fock states of the oscillator mode, with $\ket{0}$ the vacuum state.
The set of Fock states forms a complete set and the total density matrix $\rho$ can be expanded as
\begin{equation}
  \rho = \sum_{m,n = 0}^{\infty} \rho_{m,n} \otimes \ketbra{m}{n},
  \label{eq:method_rho}
\end{equation}
where $\rho_{m,n} = \braket{m|\rho|n}$ is a target system operator.
Inserting Eq.$\,$(\ref{eq:method_rho}) into Eq.$\,$(\ref{eq:method_Ltot}) yields
\begin{equation}
  \begin{gathered}
    \braket{m|\mathcal{L} (\rho)|n} = \mathcal{L}_S (\rho_{m,n}) \\
    + \sqrt{m} \, (\epsilon I_S - iV_S) \rho_{m-1,n} - \sqrt{m+1} \, (\epsilon I_S - iV_S)^\dagger \rho_{m+1,n} \\
    + \sqrt{n} \, \rho_{m,n-1} (\epsilon I_S - iV_S)^\dagger - \sqrt{n+1} \, \rho_{m,n+1} (\epsilon I_S - iV_S) \\
    + \kappa \sqrt{(m+1)(n+1)} \rho_{m+1,n+1} - \frac{\kappa}{2} (m+n) \rho_{m,n},
  \end{gathered}
  \label{eq:method_drhomn}
\end{equation}
with $I_S$ the identity operator on the target system space.
The conventional computational approach consists in integrating $(d/dt) \rho_{m,n} = \braket{m|\mathcal{L} (\rho)|n}$ with a truncation of the indices $m$ and $n$ ($\rho_{m,n} = 0$ for $m,n \geq N_{\rm tr}$).
Note that the oscillator system space after the truncation is $N_{\rm tr}$-dimensional since $m$ and $n$ include $0$.

Instead of the density matrix elements $\{ \rho_{m,n} \}$, we solve the evolution equation for quantities related to moments of the quadrature operators in this article.
In \cite{Link22}, the authors considered operators on the target system space proportional to ${\rm tr}_{\rm osc}(a^m \rho (a^\dagger)^n)$.
The formulation based on them is presented in Appendix \ref{app:RhoXiChi}.
In this work, we instead consider the following operators on the target system space,
\begin{equation}
  \chi_{m,n} = \frac{{\rm tr}_{\rm osc} \left(\Upsilon_{+}^m \Upsilon_{-}^n (\rho) \right)  }{\sqrt{m! \, n!}} \ \ \ (m,n = 0,1,2,\dots),
  \label{eq:method_chirho}
\end{equation}
with $\Upsilon_{\pm}$ defined by
\begin{equation*}
  \Upsilon_{\pm} (\rho) = a \rho \pm \rho a^\dagger.
\end{equation*}
Since we have $[\Upsilon_+, \Upsilon_-] = 0$, the order of $\Upsilon_{\pm}$ in Eq.$\,$(\ref{eq:method_chirho}) does not matter.
For later purpose, we also introduce the operator,
\begin{equation}
  \chi = \sum_{m,n = 0}^{\infty} \chi_{m,n} \otimes \ketbra{m}{n},
  \label{eq:method_chi}
\end{equation}
in the total space.
Note that the element with $m = n = 0$ of $\chi$ is the reduced density matrix of the target system $\chi_{0,0} = {\rm tr}_{\rm osc} (\rho) \equiv \rho_S$.
Accordingly, the trace of $\chi_{0,0}$ gives the trace of the total density matrix, ${\rm tr}_S (\chi_{0,0}) = {\rm tr} (\rho)$.
On the other hand, the trace of the elements $\{ \chi_{m,n} \}$ with $m + n = 1$ and $2$ read
\begin{equation}
  \begin{gathered}
    {\rm tr}_S (\chi_{1,0}) = \sqrt{2} \braket{X_{\rm osc}}, \\
    {\rm tr}_S (\chi_{0,1}) = \sqrt{2} i \braket{P_{\rm osc}}, \\
    {\rm tr}_S (\chi_{1,1}) = i \braket{ \{ X_{\rm osc}, P_{\rm osc}\} }, \\
    {\rm tr}_S (\chi_{2,0}) = (2 \braket{X_{\rm osc}^2} - 1)/\sqrt{2}, \\
    {\rm tr}_S (\chi_{0,2}) = (1 - 2 \braket{P_{\rm osc}^2})/\sqrt{2}, \\
  \end{gathered}
  \label{eq:method_LowMoments}
\end{equation}
where $\braket{O} = {\rm tr} (O \rho)$ is the expectation value of an operator $O$.
These relations indicate that ${\rm tr}_S (\chi_{m,n}) \ (m + n > 0)$ represent the moments of the quadrature operators.

It can be shown that $\{ \rho_{m,n} \}$ can be constructed by a linear combination of $\{ \chi_{m,n} \}$ (see Eq.$\,$(\ref{eq:RhoXiChi_RhoFromChi})).
In this sense, the operator $\chi$ provides a complete description of the state $\rho$.

When $\rho$ solves the master equation Eq.$\,$(\ref{eq:method_Leq}), the time evolution of $\chi_{m,n}$ is given by (see Appendix \ref{app:RhoXiChi} for the detail of this derivation)
\begin{equation}
  \frac{d}{dt} \chi_{m,n} = \braket{m|\mathcal{M}(\chi)|n},
  \label{eq:method_masterchi0}
\end{equation}
with
\begin{equation}
  \begin{gathered}
    \braket{m| \mathcal{M} (\chi) |n} = \mathcal{L}_S (\chi_{m,n}) - \frac{\kappa}{2} (m+n) \chi_{m,n} \\
    + (\epsilon+\epsilon^*) \sqrt{m} \, \chi_{m-1,n} + (\epsilon-\epsilon^*) \sqrt{n} \, \chi_{m,n-1} \\
    - i [V_S, \sqrt{m+1} \chi_{m+1,n} +  \sqrt{m} \chi_{m-1,n} ] \\
    - i \{ V_S, \sqrt{n} \chi_{m,n-1} \}.
  \end{gathered}
  \label{eq:method_dchimn}
\end{equation}
This is one of the main results of our work. We make three remarks on Eq.$\,$(\ref{eq:method_dchimn}).
\begin{enumerate}[label=(\alph*)]
  \item The reader might notice that the terms involving $\epsilon$ play different roles in the equations for $\chi_{m,n}$ (Eq.$\,$(\ref{eq:method_dchimn})) and for $\rho_{m,n}$ (Eq.$\,$(\ref{eq:method_drhomn})).
  On the one hand, in Eq.$\,$(\ref{eq:method_dchimn}), the terms involving $\epsilon$ contain only the elements with lower indices $\chi_{m-1,n}$ and $\chi_{m,n-1}$.
  On the other hand, in Eq.$\,$(\ref{eq:method_drhomn}), they contain those with not only lower indices $\rho_{m-1,n}$ and $\rho_{m,n-1}$ but also higher indices $\rho_{m+1,n}$ and $\rho_{m,n+1}$.
  The reason of this difference is explained in Appendix \ref{app:RhoXiChi} following Eq.$\,$(\ref{eq:RhoXiChi_Mchimn}).
  It is related to the fact that, in Eq.$\,$(\ref{eq:method_dchimn}), the element $\chi_{m+1,n}$ comes inside the commutator (see the third line of Eq.$\,$(\ref{eq:method_dchimn})).
  The terms that involve $\epsilon$ and $\chi_{m+1,n}$ then vanish as $\epsilon [I_S, \chi_{m+1,n}] = 0$.

  \item Note that, in Eq.$\,$(\ref{eq:method_dchimn}), $\chi_{m+1,n}$ is the only element having higher indices than $\chi_{m,n}$.
  In other words, the evolution of $\chi_{m,n}$ is independent of $\{ \chi_{p,q} \}_{p \geq 0, \ q > n}$.
  This is in stark contrast to Eq.$\,$(\ref{eq:method_drhomn}), where $\rho_{m+1,n+1}$, $\rho_{m+1,n}$, and $\rho_{m,n+1}$ appear.
  This difference can be understood as follows.
  First, the term with $\rho_{m+1,n+1}$ in Eq.$\,$(\ref{eq:method_drhomn}) results from the dissipator $\kappa \mathcal{D}[a]$.
  This dissipator is diagonalized after the transformation from $\rho$ to $\chi$ (see Eq.$\,$(\ref{eq:RhoXiChi_DiagDissipator})).
  For this reason, the term involving $\kappa$ in Eq.$\,$(\ref{eq:method_dchimn}) contains only $\chi_{m,n}$.
  Second, $\rho_{m+1,n}$ and $\rho_{m,n+1}$ result from the terms involving $\epsilon$ and the interaction term.
  The terms involving $\epsilon$ are not coupled to the higher indices in Eq.$\,$(\ref{eq:method_dchimn}) for the reason discussed in the above remark (a).
  For the terms involving $V_S$, we note from Eqs.$\,$(\ref{eq:method_LowMoments}) that the left and right indices of $\chi_{m,n}$ are associated with the position quadrature and momentum quadrature, respectively.
  Since the interaction Hamiltonian, $V_S (a + a^\dagger)$, depends only on the position quadrature,  it does not contribute to raising the right index that is associated with the momentum quadrature.
 This property in the representation of $\chi$ is advantageous in numerical simulations as explained below.
  We note that the representation proposed in \cite{Link22} (${\rm tr}_{\rm osc}(a^m \rho (a^\dagger)^n)$) does not have this property (see Eq.$\,$(\ref{eq:RhoXiChi_Mximn}) and the subsequent discussion).

  \item The evolution equation Eq.$\,$(\ref{eq:method_masterchi0}) can be solved numerically in a similar way to the equation Eq.$\,$(\ref{eq:method_drhomn}) for $\rho_{m,n}$.
  The time evolution of the reduced density matrix can then be evaluated by extracting the element with $m=n=0$, $\chi_{0,0}$.
\end{enumerate}

In situations where we seek to extract the evolution of the target quantum system, what we need is the reduced density matrix $\rho_S$, not the total density matrix $\rho$.
The solution procedure for Eq.$\,$(\ref{eq:method_masterchi0}) is then advantageous for the following reasons.
By setting $n = 0$ in Eq.$\,$(\ref{eq:method_dchimn}), we obtain
\begin{equation}
    \begin{gathered}
      \frac{d}{dt} \chi_{m,0} = \mathcal{L}_S (\chi_{m,0}) - \frac{\kappa}{2} m \chi_{m,0} + 2u \sqrt{m} \chi_{m-1,0} \\
      - i [V_S, \sqrt{m+1} \chi_{m+1,0} + \sqrt{m} \chi_{m-1,0} ],
    \end{gathered}
  \label{eq:method_dchim0}
\end{equation}
with $u = {\rm Re}(\epsilon)$ where ${\rm Re}$ denotes the real part.
This equation indicates that the evolution of $\chi_{m,0}$ is independent of $\{ \chi_{m,n} \}_{n \geq 1}$ ($m$ fixed).
Since the reduced density matrix, $\rho_S = \chi_{0,0}$, is included in the set $\{ \chi_{m,0} \}_{m \geq 0}$, the evolution of $\rho_S$ can be evaluated by solving Eq.$\,$(\ref{eq:method_dchim0}).
Using the same truncation level $N_{\rm tr}$ in the conventional approach $\{ \rho_{m,n} \}_{m,n \geq 0}$ and in the new approach $\{ \chi_{m,0} \}_{m \geq 0}$, the number of degrees of freedom in the latter simulation is reduced by a factor $N_{\rm tr}$.
This reduction enables the computation of $\rho_S$ more quickly and with less memory, especially when $N_{\rm tr}$ is large.
We discuss numerical performance in more detail in later sections.

We can rewrite Eq.$\,$(\ref{eq:method_dchim0}) in a more compact form by introducing the rectangular matrix,
\begin{equation}
  \chi^{\rm rec} = \sum_{m = 0}^\infty \chi_{m,0} \otimes \ket{m},
  \label{eq:method_barchi}
\end{equation}
the evolution of which is governed by
\begin{equation}
  \frac{d}{dt} \chi^{\rm rec} = \bar{\mathcal{M}} (\chi^{\rm rec}),
  \label{eq:method_masterchi}
\end{equation}
with
\begin{equation}
  \begin{gathered}
    \bar{\mathcal{M}} (\chi^{\rm rec}) = \mathcal{L}_S (\chi^{\rm rec}) - \frac{\kappa}{2} a^\dagger a \chi^{\rm rec} + 2u a^\dagger \chi^{\rm rec} \\
    - i [V_S, (a+a^\dagger) \chi^{\rm rec}].
  \end{gathered}
  \label{eq:method_M}
\end{equation}
When solving this equation, the initial condition of $\chi^{\rm rec}$ can be determined from that of $\rho$ using Eq.$\,$(\ref{eq:method_chirho}).
The reduced density matrix can be evaluated by extracting the zeroth element of the solution,
\begin{equation}
  \rho_S = \chi_{0,0} = \bra{0} \chi^{\rm rec}.
  \label{eq:method_rhoS}
\end{equation}

We emphasize that no approximation was made in the derivation of Eq.$\,$(\ref{eq:method_masterchi}) from Eq.$\,$(\ref{eq:method_Leq}).
In addition to this, several remarks are in order.
\begin{enumerate}[label=(\roman*)]
    \item Compared with ${\rm tr}_{\rm osc}(a^m \rho (a^\dagger)^n)$, the representation in \cite{Link22}, the primary advantage of the new representation $\chi$ emerges in the absence of the term involving $a^\dagger a$ in the oscillator Hamiltonian and with the Hermitian coupling ($L_S = L_S^\dagger$).
    As elucidated underneath Eq.$\,$(\ref{eq:RhoXiChi_Mchimn}), these are the assumptions essential for the decoupling discussed in remark (b).
    These assumptions result in the correlation function of the oscillator mode being represented by a real exponential function.
    This aligns with the scenario where the HEOM simulations are simplified as discussed in Introduction.

    \item In the derivation of Eq.$\,$(\ref{eq:method_dchim0}), we set $n = 0$ and focused only on the reduced density matrix $\rho_S$, which does not explicitly contain information of the oscillator state.
    Here we note that the use of Eq.$\,$(\ref{eq:method_dchimn}) is still advantageous when low-order moments of the quadrature operators are also needed.
    As can be inferred from Eq.$\,$(\ref{eq:method_LowMoments}), moments at order $N$ can in general be obtained from the set $\{ \chi_{m,n} \}_{m + n = N}$.
    We discussed in the remark (b) that the evolution of $\chi_{m,n}$ depends on $\{ \chi_{p,q} \}_{p \geq 0, \ 0 \leq q \leq n}$, but not on $\{ \chi_{p,q} \}_{p \geq 0, \ q > n}$.
    Accordingly, given a sufficiently large truncation level for the left index,
   one only needs to truncate the right index $n$ at $n = N + 1$ ($\chi_{m,n} = 0$ for $n \geq N+1$) to extract all moments up to order $N$.
    In this simulation, the number of degrees of freedom increases only by a factor $N + 1$ compared with solving Eq.$\,$(\ref{eq:method_masterchi}) and, for small $N$ (low-order moments), the computational cost does not change significantly.
   By a similar argument, we can also evaluate the correlation function of the quadrature operators by accounting for $\{ \chi_{m,n} \}_{m \geq 0, n = 0,1,2}$ (see discussion following Eq.$\,$(\ref{eq:OptMech_c_posc})).
    We demonstrate such a computation in Sec.$\,$\ref{sec:OptMech}.

    \item The decoupling mentioned in remark (b) is a property of the generator and independent of the initial conditions.
    Therefore, Eq.$\,$(\ref{eq:method_masterchi}) is applicable to any initial state, including non-Gaussian oscillator states and correlated states.
    For an initial state $\rho$ with $a^{N_{\rm max}} \rho = 0 \ (N_{\rm max} \geq 1)$, Eq.$\,$(\ref{eq:method_chirho}) implies that $\chi_{m, 0} = 0$ for $m \geq 2N_{\rm max} - 1$.
    This suggests that the $\chi$-representation generally requires a larger truncation level to describe an initial state.
    However, as far as the value of $N_{\rm max}$ is not excessively large, we can still expect the computational advantage.

    \item The extension to several damped oscillator modes is straightforward.

\end{enumerate}

\section{Computation of the correlation function in an optomechanical setting}
\label{sec:OptMech}

In this section, we apply the moment expansion method formulated in Sec.$\,$\ref{sec:method} to the numerical computation of the correlation function.
The purpose of this section is twofold.
One is to provide a way to calculate the correlation function within the new framework.
We rewrite the quantum regression formula using the new representation $\chi$.
The other is to demonstrate the efficiency of the new method.
By performing numerical tests, we show that the moment expansion method yields results consistent with the conventional method at a lower computational cost.

We consider an optomechanical setting \cite{RMP14}.
The system is composed of a photon mode inside a cavity and a mechanical oscillator, the annihilation operators of which are denoted by $a_c$ and $b_m$, respectively.
Here the subscripts $c$ for the cavity and $m$ for the mechanical oscillator are used to differentiate these operators from those introduced later after the linearization.
These two oscillator modes are coupled via the radiation pressure force.
Here we treat the mechanical oscillator as the target system.
We begin with the following master equation,
\begin{equation}
  \begin{gathered}
    \frac{d}{dt} \rho = - i [H_{\rm om} + H_{\rm drive}(t), \rho] \\
    + \kappa \mathcal{D}[a_c] (\rho) + \gamma (1 + n_{\rm th}) \mathcal{D} [b_m] (\rho) + \gamma n_{\rm th} \mathcal{D} [b_m^\dagger] (\rho).
  \end{gathered}
  \label{eq:OptMech_Master0}
\end{equation}
The first line of Eq.$\,$(\ref{eq:OptMech_Master0}) describes the unitary dynamics, where $H_{\rm om}$ is the optomechanical Hamiltonian given by \cite{RMP14}
\begin{equation*}
  H_{\rm om} =  \omega_c (x_m) a_c^\dagger a_c +  \omega_m b_m^\dagger b_m,
\end{equation*}
with $\omega_c$ the frequency of the relevant photon mode inside a cavity, $\omega_m$ the frequency of the mechanical oscillator,
and $x_m$ the displacement of the mechanical oscillator.
In what follows, we denote $x_m = x_{\rm zpf} (b_m + b_m^\dagger)$ with the amplitude of the zero-point fluctuation $x_{\rm zpf}$.
The photon frequency $\omega_c$ depends on the displacement $x_m$ of the mechanical oscillator.
Most optomechanical settings deal with small displacements.
For this reason, it is common to expand $\omega_c(x_m)$ around $x_m = 0$,
\begin{equation}
  \begin{gathered}
    \omega_c (x_m) = \omega_c + x_{\rm zpf} \frac{d \omega_c}{dx} (b_m + b_m^\dagger) \\
    + \frac{x_{\rm zpf}^2}{2} \frac{d^2 \omega_c}{dx^2} (b_m + b_m^\dagger)^2 + \dots,
  \end{gathered}
  \label{eq:OptMech_omegac_expand}
\end{equation}
with $\omega_c = \omega_c(x_m = 0)$, and to truncate at a low order.
Taking into account only the first-order term, the optomechanical Hamiltonian reads
\begin{equation*}
  H_{\rm om} =  \omega_c a_c^\dagger a_c +  \omega_m b_m^\dagger b_m +  g_0 a_c^\dagger a_c (b_m + b_m^\dagger),
\end{equation*}
with $g_0 = x_{\rm zpf} (d\omega_c/dx_m)_{x_m = 0}$ the bare coupling strength.
In the first line of Eq.$\,$(\ref{eq:OptMech_Master0}), $H_{\rm drive} (t)$ represents the coherent drive to the cavity and is given by $H_{\rm drive} (t) = i  (\alpha_d \, a_c^\dagger e^{-i \omega_d t} - \alpha_d^* \, a_c e^{i \omega_d t})$ with $\alpha_d$ the drive amplitude and $\omega_d$ the drive frequency.
The second line of Eq.$\,$(\ref{eq:OptMech_Master0}) describes the dissipative dynamics, where $\kappa$ and $\gamma$ are the loss rates of the photon and mechanical oscillator, respectively, and $n_{\rm th}$ is the average number of thermal quanta associated with the mechanical oscillator.

The right-hand side of Eq.$\,$(\ref{eq:OptMech_Master0}) depends explicitly on time due to the drive term $H_{\rm drive}(t)$.
This time dependence can be removed by moving to the frame rotating with the drive frequency, namely by considering $\rho_d = \exp(i \omega_d t \, a_c^\dagger a_c) \, \rho \, \exp(- i \omega_d t \, a_c^\dagger a_c)$,
\begin{equation}
  \begin{gathered}
    \frac{d}{dt} \rho_d = - i [H, \rho_d] + \kappa \mathcal{D}[a_c] (\rho_d) \\
    + \gamma (1 + n_{\rm th}) \mathcal{D} [b_m] (\rho_d) + \gamma n_{\rm th} \mathcal{D} [b_m^\dagger] (\rho_d),
  \end{gathered}
  \label{eq:OptMech_Master1}
\end{equation}
where we introduce
\begin{equation*}
  \begin{gathered}
    H = -  \Delta a_c^\dagger a_c +  \omega_m b_m^\dagger b_m \\
    -  g_0 a_c^\dagger a_c (b_m + b_m^\dagger) + i  (\alpha_d \, a_c^\dagger - \alpha_d^* \, a_c ),
  \end{gathered}
\end{equation*}
with the cavity detuning $\Delta = \omega_d - \omega_c$.

The Heisenberg equations of motion for $a_c$ and $b_m$ derived from Eq.$\,$(\ref{eq:OptMech_Master1}) are nonlinear due to the coupling term $a_c^\dagger a_c (b_m + b_m^\dagger)$.
Common experimental settings consist of a strong drive $\alpha_d$ and a weak bare coupling $g_0$.
In such cases, the master equation Eq.$\,$(\ref{eq:OptMech_Master1}) can be linearized approximately \cite{RMP14}.
The linearization is performed by dividing the annihilation operators ($a_c$, $b_m$) into the classical steady state values ($\bar{\alpha}$, $\bar{\beta}$) and the fluctuations around them ($a$, $b$) as $a_c = \bar{\alpha} + a$ and $b_m = \bar{\beta} + b$.
Note that $a$ and $b$ are ladder operators, that is, they satisfy $[a,a^\dagger] = 1$ and $[b,b^\dagger] = 1$.
Without loss of generality, we assume that $\bar{\alpha}$ is real.
In terms of $a$ and $b$, the coupling Hamiltonian reads $ g_0 (\bar{\alpha} (a + a^\dagger) + a^\dagger a) (b + b^\dagger)$.
For a weak bare coupling $g_0$, the leading order of $\bar{\alpha}$ is proportional to $\alpha_d$.
Then, for a strong drive $\alpha_d$, the term $a^\dagger a$ is smaller than the term $\bar{\alpha} (a + a^\dagger) $ by a factor $|\bar{\alpha}|$ and can be neglected approximately.
These considerations lead to the linearized master equation,
\begin{equation}
  \begin{gathered}
    \frac{d}{dt} \rho_d = - i [H_{\rm lin}, \rho_d] \\
    + \kappa \mathcal{D}[a] (\rho_d) + \gamma (1 + n_{\rm th}) \mathcal{D} [b] (\rho_d) + \gamma n_{\rm th} \mathcal{D} [b^\dagger] (\rho_d),
  \end{gathered}
  \label{eq:OptMech_Master}
\end{equation}
with the linearized Hamiltonian,
\begin{equation}
  H_{\rm lin} = -  \Delta' a^\dagger a +  \omega_m b^\dagger b +  g (a + a^\dagger)(b + b^\dagger),
  \label{eq:OptMech_Hlin}
\end{equation}
with $\Delta' = \Delta + 2 g_0 {\rm Re}(\bar{\beta})$ and $g = g_0 \bar{\alpha}$.
In the following, we assume that $\omega_d$ is set so that $\Delta' = 0$.
Note that Eq.$\,$(\ref{eq:OptMech_Master}) has the form of Eq.$\,$(\ref{eq:method_Ltot}) with $V_S = g (b + b^\dagger)$,
$\mathcal{L}_S(\rho) = - i \omega_m [ b^\dagger b, \rho] + \gamma (1 + n_{\rm th}) \mathcal{D} [b] (\rho) + \gamma n_{\rm th} \mathcal{D} [b^\dagger] (\rho)$, and $\epsilon = 0$.

In Eq.$\,$(\ref{eq:OptMech_Master0}), we have used the master equation approach to take into account the dissipative dynamics.
Another approach, which is often employed when studying optomechanical systems \cite{RMP14}, is the quantum Langevin equations \cite{QuantumNoise}.
In the present context, they are the equations of motion for $a_c$ and $b_m$ including the operator-valued white noise associated with environment degrees of freedom.
The master equation of the form Eq.$\,$(\ref{eq:OptMech_Master0}) can be derived after tracing out environment degrees of freedom \cite{QuantumNoise}.
For the linearized Hamiltonian Eq.$\,$(\ref{eq:OptMech_Hlin}), the quantum Langevin equations are linear with respect to $a_c$ and $b_m$ and can be solved analytically.
When a nonlinear term is present, on the other hand, solving the quantum Langevin equations is not straightforward, even numerically.
In such cases, the master equation approach provides a more powerful tool.
We consider a nonlinear coupling below (see discussions following Eq.$\,$(\ref{eq:OptMech_quad})).

For numerical tests, we compute the correlation function of the quadrature operators of the photon and the mechanical oscillator.
In optomechanical settings, it can be extracted experimentally using homodyne detection.
The correlation function can be used for various purposes.
For instance, the power spectrum, which is given by the Fourier transform of the correlation function, can reveal detail of the coupling \cite{Brawley16} as will be demonstrated in a simple setting below (see Fig.$\,$\ref{fig:OptMech_corrspec_lin_quad} and discussions in the text).

Let us recall how to calculate the correlation function in the conventional method.
The discussions in this and the next paragraphs are not restricted to optomechanical systems, and we consider general settings.
The correlation function is defined including environment degrees of freedom.
In this discussion, the target plus oscillator system is referred to as the total system following the previous terminology, and the total system plus environment is referred to as the universe.
The correlation function of a total system operator $O$ is defined by
\begin{equation*}
  C_O(t) = {\rm tr}_{\rm unv} (e^{i H_{\rm unv} t} \Delta O e^{- i H_{\rm unv} t} \Delta O \rho^{\rm unv}_{\rm eq}),
\end{equation*}
where ${\rm tr}_{\rm unv}$ is the trace over the universe, $H_{\rm unv}$ is a Hamiltonian of the universe, $\rho^{\rm unv}_{\rm eq}$ is an equilibrium density operator on the universe satisfying $[\rho^{\rm unv}_{\rm eq}, H_{\rm unv}] = 0$, and $\Delta O = O - {\rm tr}_{\rm unv}(O \rho^{\rm unv}_{\rm eq})$.
By inserting the definition of $\Delta O$, we obtain
\begin{equation*}
  \begin{gathered}
    C_O(t) = {\rm tr}_{\rm unv} (O e^{- i H_{\rm unv} t} (O \rho^{\rm unv}_{\rm eq}) e^{i H_{\rm unv} t}) \\ - [{\rm tr}_{\rm unv}(O \rho^{\rm unv}_{\rm eq})]^2.
  \end{gathered}
\end{equation*}
With the Markov approximation, we can use the quantum regression formula \cite{QuantumNoise} to calculate the correlation function from the solution $\rho$ to a master equation.
For a total system operator $O$, it reads
\begin{equation}
  C_O(t) = {\rm tr} (O e^{\mathcal{L} t}( O \rho_{ss})) - [{\rm tr} (O \rho_{ss})]^2,
  \label{eq:OptMech_c_o}
\end{equation}
where $\rho_{ss}$ is a steady state solution to the master equation.
The first term on the right-hand side can be evaluated as follows.
For an initial state $\rho$, we first solve the master equation Eq.$\,$(\ref{eq:method_Leq}) until the solution reaches a steady state, $\rho_{ss} = \lim_{t \to \infty} e^{\mathcal{L} t} (\rho)$.
With $\rho_{ss}$, we next solve the master equation with the initial condition $O \rho_{ss}$ up to time $t$.
This gives $e^{\mathcal{L} t}( O \rho_{ss})$ in the above formula.
Then the expectation value of $O$ with respect to the solution gives the first term.
Once the correlation function is obtained, we can evaluate the power spectrum $S_O(\omega)$ by the Fourier transform of it as
\begin{equation}
  S_O(\omega) = {\rm Re} \int_0^\infty dt \, e^{i \omega t} C_O(t).
  \label{eq:OptMech_s_o}
\end{equation}

We next rewrite the formula Eq.$\,$(\ref{eq:OptMech_c_o}) using the new representation $\chi$.
Detailed calculations are presented in Appendix \ref{app:RhoXiChi}, and here we summarize the results.
When an operator $O$ acts only on the target system space, the correlation function reads (see Eq.$\,$(\ref{eq:RhoXiChi_autocorr_sys}))
\begin{equation*}
  C_O(t) = {\rm tr}_S (O \braket{0| e^{\mathcal{M} t} ( O \chi_{ss} )|0}) - [{\rm tr}_S (O \braket{0|\chi_{ss}|0} )]^2,
\end{equation*}
where $\chi_{ss}$ is a steady-state solution to Eq.$\,$(\ref{eq:method_masterchi0}).
Note that the state of the oscillator mode is not affected by the operation of $O$.
In addition, we need only the element with $m = n = 0$ of $\chi$ eventually.
Therefore, by the same argument as that underneath Eq.$\,$(\ref{eq:method_dchim0}), the correlation function can be evaluated with the rectangular matrix and the corresponding generator as
\begin{equation*}
  C_O(t) = {\rm tr}_S (O \bra{0} e^{\bar{\mathcal{M}} t} ( O \chi^{\rm rec}_{ss} ) ) - [{\rm tr}_S (O \bra{0} \chi^{\rm rec}_{ss} )]^2.
\end{equation*}
When $O$ is the momentum quadrature operator $P_{\rm osc}$ of the photon mode, on the other hand, the correlation function reads (see Eq.$\,$(\ref{eq:RhoXiChi_autocorr_p_osc}))
\begin{equation}
  \begin{gathered}
    C_{P_{\rm osc}} (t) = \frac{i}{\sqrt{2}} {\rm tr}_S \braket{0| e^{\mathcal{M} t} ( \chi_{ss} P_{\rm osc}- \frac{i}{\sqrt{2}} a^\dagger \chi_{ss} ) |1} \\
    + \frac{1}{2} [{\rm tr}_S \braket{0|\chi_{ss}|1} ]^2.
    \label{eq:OptMech_c_posc}
  \end{gathered}
\end{equation}
Since $P_{\rm osc}$ is operated from the right side and the $m = 0, \ n = 1$ element is extracted at the end, the calculation of $C_{P_{\rm osc}} (t)$ requires only the elements $\{ \chi_{m,n} \}_{m \geq 0, n = 0,1,2}$.

To demonstrate the performance of this approach, we compute the correlation functions using the conventional method ($\rho$) and the moment expansion method ($\chi$).
Denoting the position quadrature operator of the mechanical oscillator by $X_{\rm mec} = (b + b^\dagger)/\sqrt{2}$, we compute $C_{X_{\rm mec}}(t)$ and $C_{P_{\rm osc}}(t)$.
We use the Fock bases both for the photon and the mechanical oscillator states and denote their truncation levels by $N_{\rm osc}$ and $N_{\rm mec}$, respectively.
We recall three points regarding the truncation.
First, the truncation level is said to be $N_{\rm osc}$ when the Fock states $\{ \ket{n} \}_{0 \leq n \leq N_{\rm osc}-1}$ are taken into account.
Second, in the moment expansion method, we track the evolution of the elements $\{ \chi_{m,n} \}_{0 \leq m \leq N_{\rm osc}-1, \, 0 \leq n \leq N_r-1}$ with $N_r = 1$ for the computation of $C_{X_{\rm mec}}(t)$ and $N_r = 3$ for that of $C_{P_{\rm osc}}(t)$.
Third, the number of degrees of freedom reads $N_{\rm osc}^2 N_{\rm mec}^2$ for $\rho$ and $N_{\rm osc} N_rN_{\rm mec}^2$ for $\chi$, and the latter is smaller than the former by a factor $N_{\rm osc}/N_r$.
When computing the steady state, the initial condition is assumed to be
\begin{equation*}
  \rho(t=0) = \frac{ e^{-\beta  \omega_m b^\dagger b} }{{\rm tr}_S (e^{-\beta  \omega_m b^\dagger b}) } \otimes \ketbra{0}{0},
\end{equation*}
which leads to
\begin{equation*}
  \chi(t=0) = \frac{ e^{-\beta  \omega_m b^\dagger b} }{{\rm tr}_S (e^{-\beta  \omega_m b^\dagger b}) } \otimes \ketbra{0}{0},
\end{equation*}
where the inverse temperature of the environment $\beta$ is determined from $e^{\beta  \omega_m} = 1 + n_{\rm th}^{-1}$ with $n_{\rm th}$ in Eq.$\,$(\ref{eq:OptMech_Master}).
In our simulations, we set $n_{\rm th} = 1$.
In the unit of time such that $\kappa = 1$, we arbitrarily set $\omega_m = 1$, $\gamma = 0.1$ and $g = 0.25$.
For numerical integration of the master equations, we employ the fourth-order Runge-Kutta method with the time step $\Delta t = 0.02$.
The steady states are obtained by $\rho_{ss} = e^{\mathcal{L} t_{ss}} (\rho)$ and $\chi_{ss} = e^{\mathcal{M} t_{ss}} (\chi)$ with the above initial conditions.
Through numerical experiments, we found that a value of $t_{ss} = 50$ suffices to achieve a steady state, and this value is  employed in subsequent simulations.
All the computations in this article are performed using Python on a MacBook Air (M1, 2020) equipped with Apple M1 chip, and double precision floating-point numbers are employed unless specified.

The truncation levels $N_{\rm osc}$ and $N_{\rm mec}$ are determined following Appendix \ref{app:det.trunc}.
With $\vec{N} = (N_{\rm osc}, N_{\rm mec})$, we obtained as a result $\vec{N}_\rho = (12,30)$ in the conventional method $(\rho)$ and $\vec{N}_\chi = (3,24)$ in the moment expansion method $(\chi)$.
The result $N_{\rm osc} = 3$ for $\chi$ is expected because we now consider the moments equal to or lower than the second-order and their evolution is independent of the higher-order moments in the linearized master equation Eq.$\,$(\ref{eq:OptMech_Master}).
The smaller value of $N_{\rm mec}$ in the moment expansion method would have no special significance.
In the second application presented below, the two methods yield the same value of $N_{\rm mec}$.

\begin{figure}[t]
  \centering
  \includegraphics[keepaspectratio, scale=0.5]{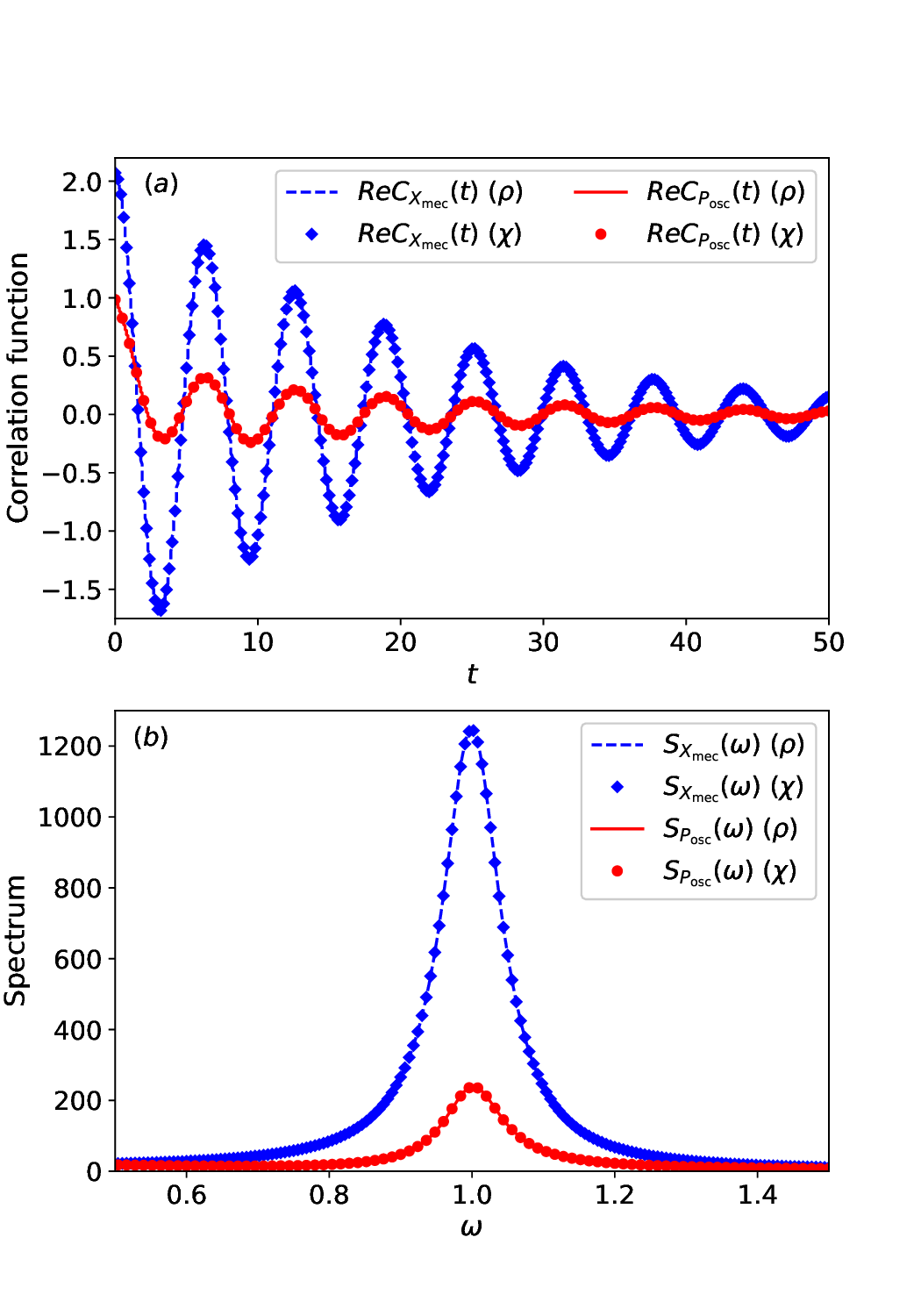}
  \caption{Correlation functions and power spectra of $X_{\rm mec}$ (dashed blue lines and blue diamonds) and $P_{\rm osc}$ (solid red lines and red circles) computed with the conventional method ($\rho$) and the moment expansion method ($\chi$). The time $t$, the frequency $\omega$, and the spectrum are shown in the unit of $\kappa = 1$.
  (a) The real part of the correlation functions defined by Eq.$\,$(\ref{eq:OptMech_c_o}).
  (b) The power spectrum defined by Eq.$\,$(\ref{eq:OptMech_s_o}).
  }
  \label{fig:OptMech_corrspec_lin}
\end{figure}

Fig.$\,$\ref{fig:OptMech_corrspec_lin} $(a)$ shows the resulting correlation functions.
We see that the results of the moment expansion method, shown by the blue diamonds and red circles, agree with those of the conventional method, shown by the dashed blue and solid red lines
(a quantitative comparison will be made below using power spectrum results).
As shown in the first and second columns of Table \ref{tab:OptMech_comptime_lin}, the computation of $C_{X_{\rm mec}}(t)$ with the moment expansion method can be performed about 82 times faster than with the conventional method.
The computation of $C_{P_{\rm osc}}(t)$ can also be sped up by a factor 28 (see the third column of Table \ref{tab:OptMech_comptime_lin}).
We stress that the faster computation is not only due to the smaller truncation levels.
The last column of Table \ref{tab:OptMech_comptime_lin} shows the computational time for the moment expansion method with the same truncation levels as the conventional method.
We achieve a speedup by a factor 4, highlighting the efficiency of the new method.
This speedup factor can be attributed to the reduction in the number of degrees of freedom in the simulation of $\chi$.
As previously mentioned, this reduction is quantified by $N_{\rm osc}/N_r = 12/3 = 4$ in the current setting.

We also compute the power spectrum from the correlation function.
The time integral in Eq.$\,$(\ref{eq:OptMech_s_o}) is performed up to $t = 150$ at which the amplitude of the correlation functions becomes in the order of $10^{-4}$.
The result is shown in Fig.$\,$\ref{fig:OptMech_corrspec_lin} $(b)$.
Even though the power spectrum is sensitive to a small change in the correlation function, we see agreement between the two methods.
The relative errors, $|S_O^\rho(\omega)-S_O^\chi(\omega)|/|S_O^\rho(\omega)|$ with $S_O^\rho(\omega)$ ($S_O^\chi(\omega)$) the spectrum computed using $\rho$ ($\chi$), in the region shown in Fig.$\,$\ref{fig:OptMech_corrspec_lin} $(b)$ are in the order of $10^{-4}$.

\begin{table}[h]
  \centering
  \begin{tabular}{|c|c|c|c|c|} \hline
    & $ \rho_{m,n} \, (\vec{N}_\rho) $ & $ \chi_{m,0} \, (\vec{N}_\chi) $ & $ \chi_{m,n \leq 2} \, (\vec{N}_\chi) $ & $ \chi_{m,n \leq 2} \, (\vec{N}_\rho) $ \\ \hline
   time (sec) & 2.97 & 3.58 $\times 10^{-2}$ & 1.04 $\times 10^{-1}$ & 6.60 $\times 10^{-1}$ \\  \hline
  \end{tabular}
  \caption{
  Computational times for the 100 time steps in the different methods.
  In parentheses, $\vec{N}_\rho$ and $\vec{N}_\chi$ are the employed truncation levels in the computation, where $\vec{N}_\rho = (N_{\rm osc} = 12, N_{\rm mec} = 30)$ and $\vec{N}_\chi = (3,24)$.
  }
  \label{tab:OptMech_comptime_lin}
\end{table}

In the above simulations, the Liouvillian is quadratic in the ladder operators, $a$ and $b$, and is analytically solvable.
As a more nontrivial example, let us take into account the second-order term in the expansion formula Eq.$\,$(\ref{eq:OptMech_omegac_expand}) for $\omega_c(x_m)$.
Similarly to the above case, we displace the annihilation operators $a_c$ and $b_m$ and neglect the interaction term involving $a^\dagger a$.
Following \cite{Hauer18}, we consider the following master equation
\begin{equation*}
  \begin{gathered}
    \frac{d}{dt} \rho_d = - i [\omega_m b^\dagger b + (a+a^\dagger) V_S, \rho_d] \\
    + \kappa \mathcal{D}[a] (\rho_d) + \gamma (1 + n_{\rm th}) \mathcal{D} [b] (\rho_d) + \gamma n_{\rm th} \mathcal{D} [b^\dagger] (\rho_d),
  \end{gathered}
\end{equation*}
with
\begin{equation}
  V_S = g_{\rm lin} (b + b^\dagger) + g_{\rm quad} (2 b^\dagger b + b^2 + (b^\dagger)^2).
  \label{eq:OptMech_quad}
\end{equation}
In most optomechanical settings, the higher-order terms are negligible.
In the so called membrane-in-the-middle system where a dielectric membrane is placed inside the cavity, on the other hand,
the strength of the higher-order terms can be controlled by changing the location of a membrane \cite{Thompson08,Sankey10}.
Inclusion of the second-order term attracts attention for the following reason.
When a membrane is located at a node of the oscillator mode, one can realize a situation such that $g_{\rm lin} = 0$ and $g_{\rm quad} \ne 0$ in Eq.$\,$(\ref{eq:OptMech_quad}).
In this case, after eliminating the terms that involve $b^2$ and $(b^\dagger)^2$ in Eq.$\,$(\ref{eq:OptMech_quad}) using the rotating wave approximation, the Hamiltonian part commutes with the number operator $b^\dagger b$.
Then, the act of measuring the number operator $b^\dagger b$, which is related to the membrane's mechanical energy, does not disturb the quantum state of the membrane \cite{QuantumMeasurement}.
In other words, such a coupling enables a quantum nondemolition measurement of the membrane's mechanical energy, which paves the way to observe the quantized energy of a macroscopic object.

\begin{figure}[t]
  \includegraphics[keepaspectratio, scale=0.5]{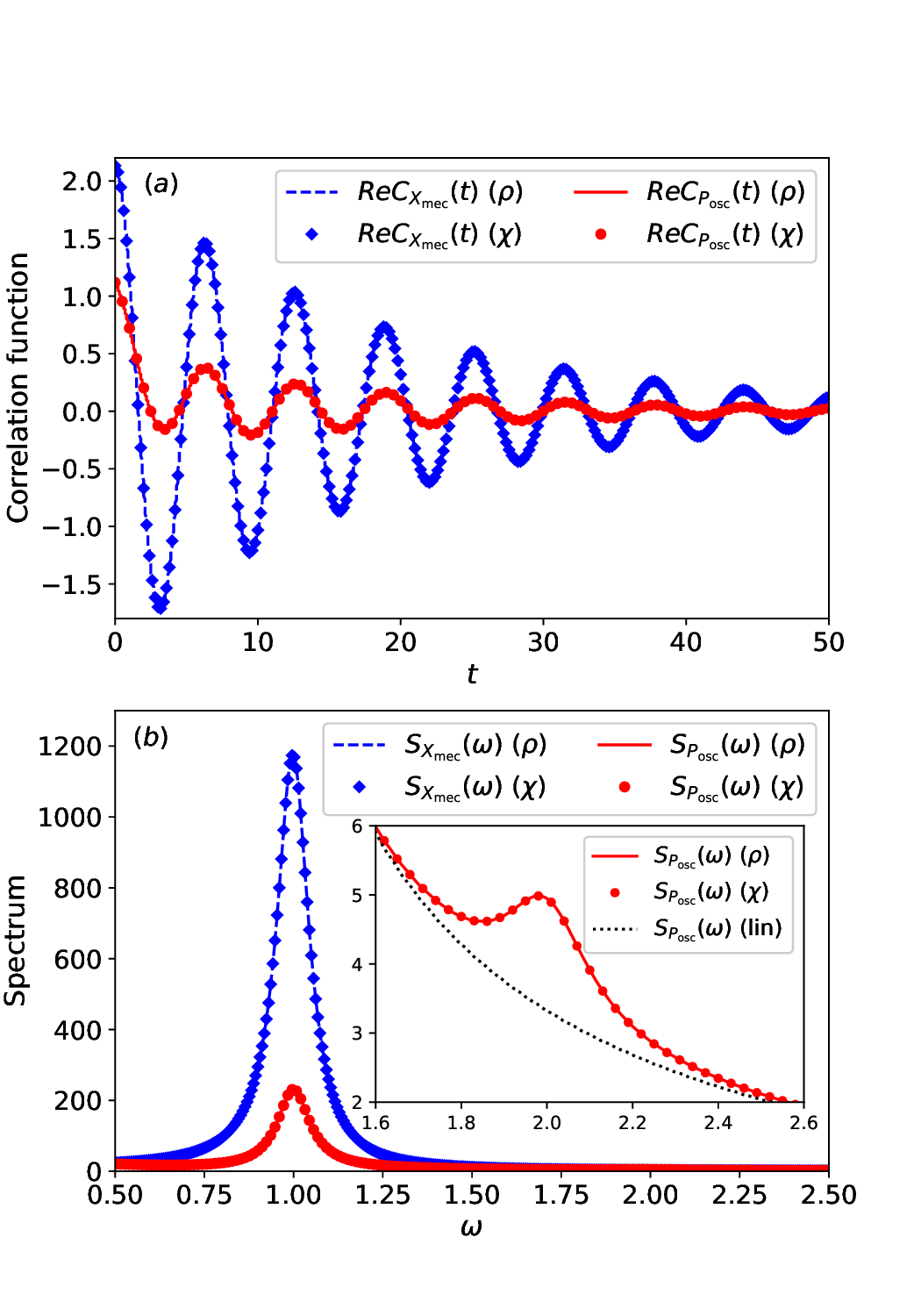}
  \caption{
  Numerical results analogous to those of Fig.$\,$\ref{fig:OptMech_corrspec_lin} in the case of the quadratic interaction given by Eq.$\,$(\ref{eq:OptMech_quad}).
  The inset of the figure $(b)$ is a zoom for an argument between $\omega = 1.6$ and $\omega = 2.6$, and the dotted black line shows the result of $S_{P_{\rm osc}}(\omega)$ without the quadratic coupling (the solid red line in Fig.$\,$\ref{fig:OptMech_corrspec_lin} $(b)$).
  }
  \label{fig:OptMech_corrspec_lin_quad}
\end{figure}

We keep both the $g_{\rm lin}$ and $g_{\rm quad}$ terms in Eq.$\,$(\ref{eq:OptMech_quad}) as in \cite{Hauer18} and compute $C_{X_{\rm mec}}(t)$ and $C_{P_{\rm osc}}(t)$.
To compare with the previous simulation (Fig.$\,$\ref{fig:OptMech_corrspec_lin}), we set $g_{\rm lin} = 0.25$ and $g_{\rm quad} = g_{\rm lin}/10$.
The other parameters remain identical.
The truncation levels are determined as in the previous simulation and read $\vec{N}_\rho = (36,30)$ in the conventional method and $\vec{N}_\chi = (6,30)$ in the moment expansion method.
Figure \ref{fig:OptMech_corrspec_lin_quad} $(a)$ shows the resulting correlation functions.
We see that the moment expansion method reproduces the result of the conventional method.
The computational times are listed in Table \ref{tab:OptMech_comptime_lin_quad}.
In this simulation, the computation of $C_{P_{\rm osc}}(t)$ with the moment expansion method can be performed about 84 times faster than with the conventional method (see the first and third columns of Table \ref{tab:OptMech_comptime_lin_quad}).
Even when $\vec{N}_\rho$ is used in the moment expansion method (the last column of Table \ref{tab:OptMech_comptime_lin_quad}), we gain a speedup by a factor 13.
This factor is consistent with the reduced number of degrees of freedom in the simulation of $\chi$, which reads $N_{\rm osc}/N_r = 36/3 = 12$ in the current setting.
We also compute the power spectrum as in the previous simulation.
The results are shown in Fig.$\,$\ref{fig:OptMech_corrspec_lin_quad} $(b)$,
the inset of which shows a zoom around $\omega = 2$.
Due to the $g_{\rm quad}$ term, $S_{P_{\rm osc}}(\omega)$ exhibits a second peak at $\omega = 2$, which is not present when $g_{\rm quad} = 0$ as can be seen from the dotted black line in the figure.
While the height of the peak is about 100 times smaller than the main peak at $\omega = 1$, the moment expansion method succeeds in capturing such fine structure.
In fact, the relative errors in the shown region are in the order of $10^{-4}$.

\begin{table}[h]
  \centering
  \begin{tabular}{|c|c|c|c|c|} \hline
    & $ \rho_{m,n} \, (\vec{N}_\rho) $ & $ \chi_{m,0} \, (\vec{N}_\chi) $ & $ \chi_{m,n \leq 2} \, (\vec{N}_\chi) $ & $ \chi_{m,n \leq 2} \, (\vec{N}_\rho) $ \\ \hline
   time (sec) & 27.4 & 1.06 $\times 10^{-1}$ & 3.24 $\times 10^{-1}$ & 1.98 \\  \hline
  \end{tabular}
  \caption{
  Computational times for the 100 time steps in the different methods.
  The employed truncation levels in each computation are $\vec{N}_\rho = (N_{\rm osc} = 36, N_{\rm mec} = 30)$ and $\vec{N}_\chi = (6,30)$.
  }
  \label{tab:OptMech_comptime_lin_quad}
\end{table}

So far, we have assumed that the oscillator is initially in the ground state, which is a Gaussian state.
As noted in remark (iii) of Sec.$\,$\ref{sec:method},
the moment expansion method is applicable to any initial state.
We conducted similar numerical tests with non-Gaussian initial oscillator states proportional to $(1-\delta) \ket{0} + \delta \ket{1} \ (0 < \delta \leq 1)$ and obtained similar results to the above, confirming the advantage of the moment expansion method.


\section{Analysis of numerical accuracy}
\label{sec:NumAccuracy}

In this section, we further investigate numerical accuracy of the moment expansion method.
As emphasized in Sec.$\,$\ref{sec:method}, the formulation of the moment expansion method itself is exact.
Numerical simulations, on the other hand, require approximations in representing real numbers, integrating the differential equation, and describing the infinite-dimensional oscillator state.
We discuss how those approximations affect numerical accuracy.
In the previous section, we discussed the numerical accuracy of the moment expansion method by comparing its results with those obtained from the conventional method, which also involve numerical errors.
In this section, we delve deeper into the accuracy of the new method by comparing it with an exact solution.

Using an exactly solvable system, we evaluate errors in the reduced density matrix computation.
As a result, it is found in Sec.$\,$\ref{subsec:NumAccuracy_plateau} that, while errors can be reduced to the order of machine precision with a sufficiently large truncation level when the parameter $u$ in Eq.$\,$(\ref{eq:method_dchim0}) is zero, simulations with nonzero $u$ exhibit a plateau in numerical accuracy.
In Sec.$\,$\ref{subsec:NumAccuracy_position}, we show that such a plateau problem can be mitigated by using the position basis.

\subsection{Plateau in numerical accuracy}
\label{subsec:NumAccuracy_plateau}

In this section, we consider a target two-level system with $\mathcal{L}_S = 0$ for the following reasons.
In order to evaluate numerical errors, we need a system that is exactly solvable.
The linearized master equation Eq.$\,$(\ref{eq:OptMech_Master}) in the previous section can be solved exactly.
However, truncation of a state space is required not only for the photon mode (the oscillator system), but also for the mechanical oscillator mode (the target system).
To avoid errors due to truncation in the target system itself, we consider a finite-dimensional target system.
If we assume $\mathcal{L}_S = 0$, the exact solution is available for any finite-dimensional target systems, as expected because $V_S$ is the only system operator in the Lindbladian in this case (see Appendix \ref{app:LDiag} for more details).
Among finite-dimensional systems, the simplest one is a two-level system.
Thus, a target two-level system is suitable for analysis of the numerical accuracy.

Let $V_S = g \sigma_x$ with $g$ the coupling constant and $\sigma_i \ (i = x,y,z)$ the Pauli matrices.
We assume that the oscillator mode is initially in the vacuum state as $\chi^{\rm rec} (t = 0) = \rho_S (0) \otimes \ket{0}$ with $\rho_S (0)$ the initial state of the two-level system.
The exact time evolution of the reduced density matrix then reads (see Eq.$\,$(\ref{eq:LDiag_ExactRhoS}))
\begin{equation}
  \begin{gathered}
    \rho_S^{\rm exact} (t) = \\
    \sum_{\alpha,\beta = \pm 1} e^{z_{\alpha,\beta} (1 - \kappa t /2 - e^{- \kappa t /2})} \ \ketbra{\alpha}{\alpha} \rho_S (0) \ketbra{\beta}{\beta},
  \end{gathered}
  \label{eq:NumAccuracy_ExactRhoS}
\end{equation}
where $\ket{\pm1} = (1/\sqrt{2}) (\pm1 \ 1)^\top$, with $\top$ the matrix transpose, is the eigenstates of $\sigma_x$ and $z_{\alpha,\beta} = 4 (\alpha - \beta) ((\alpha - \beta) + 2iu)/\kappa^2$.

Throughout this section, we present the results starting with the initial two-level state $\rho_S(0)$ as
\begin{equation}
    \rho_S(0) = \frac{1}{2} (I_2 - \sigma_z),
    \label{eq:NumAccuracy_Initialtwo-level}
\end{equation}
with $I_2$ the two-dimensional identity matrix
The dependence of numerical accuracy on $\rho_S(0)$ will be discussed in each numerical test.
To obtain the time evolution of the reduced density matrix, we integrate the evolution equation Eq.$\,$(\ref{eq:method_masterchi}) for $\chi^{\rm rec}$ and then extract its zeroth element (see Eq.$\,$(\ref{eq:method_rhoS})).
To avoid the discretization error of time $t$, we directly compute $\chi^{\rm rec}(t) = \exp(\bar{\mathcal{M}}t) (\chi^{\rm rec} (t=0))$.
For this computation, we vectorize the rectangular matrix as $\chi^{\rm rec} = \ketbra{target_1}{target_2} \otimes \ket{osc} \to \vec{\chi}^{\rm \, rec} = \ket{target_2} \otimes \ket{target_1} \otimes \ket{osc}$ and compute the matrix exponentiation of the generator $\bar{\mathcal{M}}$ using the SciPy function ${\rm scipy.linalg.expm}$.
This integration method was avoided in the previous section because the number of degrees of freedom in the conventional method was so large that the matrix exponentiation was unfeasible within our computational resource.
To estimate the numerical errors in the computed reduced density matrix $\rho_S^{\rm comp}$, we use the Frobenius norm $\lVert \rho_S^{\rm comp} - \rho_S^{\rm exact} \rVert_F = \sqrt{{\rm tr}_S [(\rho_S^{\rm comp} - \rho_S^{\rm exact})^\dagger (\rho_S^{\rm comp} - \rho_S^{\rm exact})] }$.

\begin{figure*}[t]
 \centering
  \includegraphics[keepaspectratio, scale=0.7]{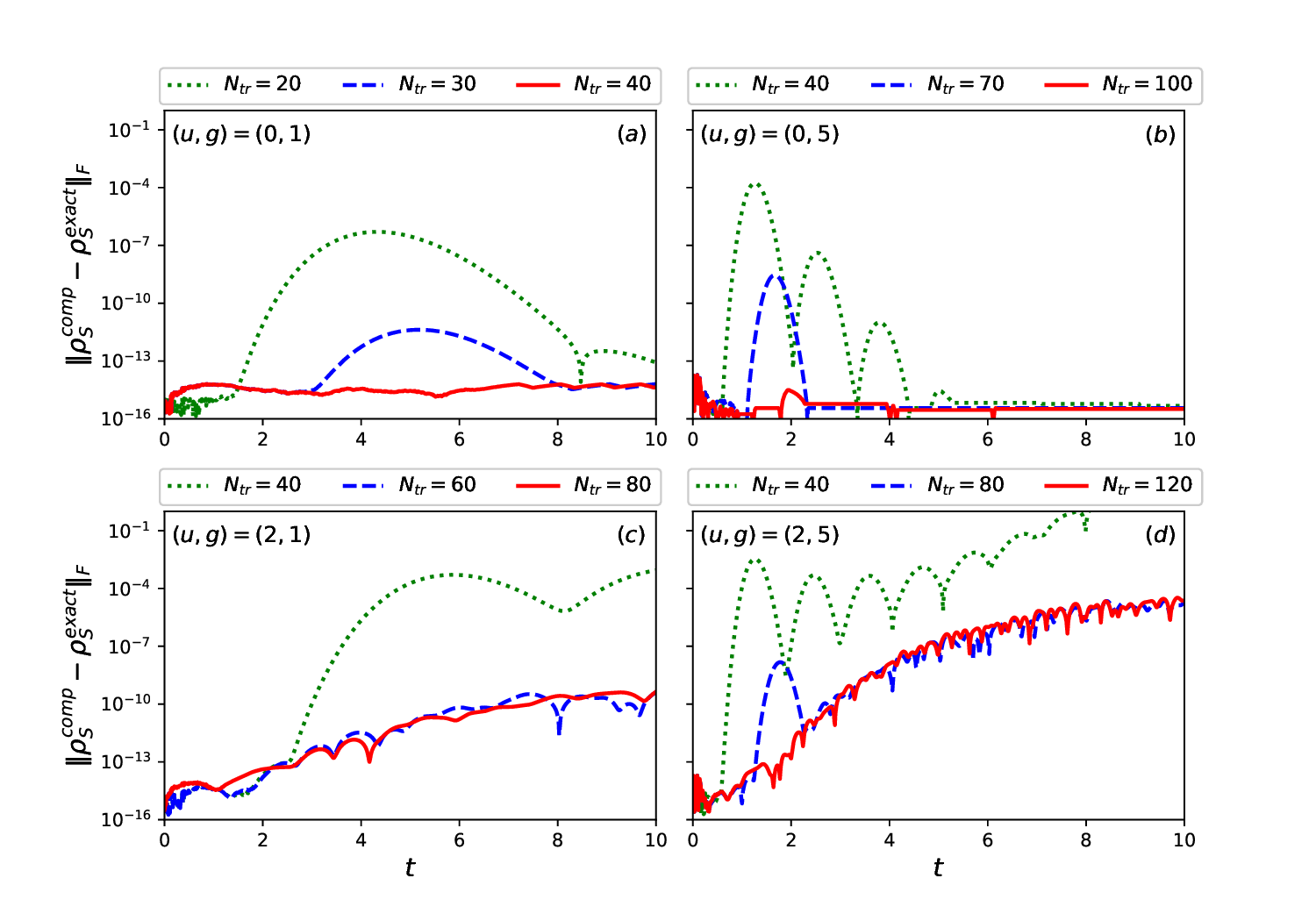}
  \caption{
  Dependence of the numerical errors on the truncation level $N_{\rm tr}$ and the values of the parameters $u$ and $g$ (we fix $\kappa = 1$).
  The time $t$ is shown in the unit of $\kappa = 1$.
  The figures in the same column have the same value of $g$ while the figures in the same row share the same value of $u$.
  The initial two-level state is given by Eq.$\,$(\ref{eq:NumAccuracy_Initialtwo-level}).
  }
  \label{fig:NumAccuracy_errors_ugdep}
\end{figure*}

In the time unit such that $\kappa = 1$, we vary the values of the parameters $u$ and $g$.
The resulting numerical errors obtained with different truncation levels $N_{\rm tr}$ are shown in Fig.$\,$\ref{fig:NumAccuracy_errors_ugdep}.
In Figs.$\,$\ref{fig:NumAccuracy_errors_ugdep} $(a)$ and $(b)$ (the first row), we set $u = 0$.
In these cases, we see that sufficiently large truncation levels ($N_{\rm tr} = 40$ in Fig.$\,$\ref{fig:NumAccuracy_errors_ugdep} $(a)$ and $N_{\rm tr} = 100$ in Fig.$\,$\ref{fig:NumAccuracy_errors_ugdep} $(b)$) allow us to reach the numerical errors in the order of machine precision ($10^{-16} - 10^{-14}$).
The same behavior is confirmed for larger values of $g$ (we checked up to $g = 20$).

In Figs.$\,$\ref{fig:NumAccuracy_errors_ugdep} $(c)$ and $(d)$ (the second row), we set $u = 2$.
In contrast with the cases with $u = 0$, we see lower bounds on the numerical error in the long-time domain.
In Fig.$\,$\ref{fig:NumAccuracy_errors_ugdep} $(c)$, the numerical errors in the long-time domain cannot be reduced less than $10^{-10}$.
In Fig.$\,$\ref{fig:NumAccuracy_errors_ugdep} $(d)$, $\rho_S^{\rm comp}$ with the smallest truncation level $N_{\rm tr} = 40$ (the dotted green line) diverges in the long-time domain.
In this case, we detected the positivity violation of $\rho_S^{\rm comp}$, while its trace is preserved.
Although such instability can be resolved by increasing the truncation level, the numerical error cannot be less than $10^{-5}$ in the long-time domain.
When the value of $u$ is further increased, the divergence of $\rho_S^{\rm comp}$ is observed even with a large truncation level.
For instance, when $(u,g) = (4,5)$, we observed the divergence even with $N_{\rm tr} = 2000$.
As in the above case, this divergence occurs in conjunction with the positivity violation of $\rho_S^{\rm comp}$.

To summarize, when $u$ is nonzero, the numerical accuracy in the moment expansion method plateaus above a truncation level.
In Appendix \ref{app:plateau_detail}, we examine the plateau problem in detail.
We find that the plateau remains even with much larger truncation levels than the levels employed in Figs.$\,$\ref{fig:NumAccuracy_errors_ugdep} $(c)$ and $(d)$.
We also find that the plateau appears independent of the initial two-level state $\rho_S(0)$.

To make the moment expansion method applicable to the largest possible set of parameter regimes, it is crucial to tackle the plateau observed in numerical accuracy.
Our investigations have identified two ways to mitigate this challenge.
The first is to increase machine precision in simulations.
In Appendix \ref{app:plateau_mp}, we show a reduction in numerical errors in the long-time domain when employing the extended precision floating-point number, compared with the double precision floating-point number as Fig.$\,$\ref{fig:NumAccuracy_errors_ugdep}.
This indicates that the errors in the long-time domain stem from roundoff errors.
Such errors are peculiar to the cases where the plateau exists; we confirmed that the difference between the double and extended precision results is insignificant (in the order of $10^{-15}$) for both the conventional method and the moment expansion method with $u = 0$, neither of which exhibits the plateau.
The peculiarity of the term involving $u$ in Eq.$\,$(\ref{eq:method_M}) is investigated in Appendix \ref{app:cond}.
This analysis reveals an inherent numerical sensitivity that has not been recognized in related research.
The other way to deal with the plateau is to use the eigenbasis of the position quadrature operator.
This is more practical than increasing machine precision and is discussed in Sec.$\,$\ref{subsec:NumAccuracy_position}.

\subsection{Eigenbasis of the position quadrature operator}
\label{subsec:NumAccuracy_position}

So far, we have used the Fock basis to represent the oscillator state.
In this subsection, we consider the eigenbasis of the position quadrature operator $X_{\rm osc} = (a + a^\dagger)/\sqrt{2}$ instead.
The use of such basis (hereafter referred to as the position basis) is motivated by the work in \cite{TA22}, where the authors studied the strong-coupling regime using HEOM and found that the position basis allowed for more stable and efficient results than the Fock basis.

The position basis vectors, $\{ \ket{x} \}_{- \infty \leq x \leq \infty}$ with $X_{\rm osc} \ket{x} = x \ket{x}$ satisfy
the orthonormality $\braket{x|x'} = \delta(x-x')$ with the Dirac's delta function $\delta(x)$ and the completeness relation $\int_{-\infty}^\infty dx \, \ketbra{x}{x} = I_{\rm osc}$ with $I_{\rm osc}$ the identity operator on the oscillator system space.
Accordingly, $\chi^{\rm rec}$ can be expanded as
\begin{equation*}
  \chi^{\rm rec} = \int_{-\infty}^\infty dx \, \chi_x \otimes \ket{x},
\end{equation*}
with $\chi_x = \bra{x} \chi^{\rm rec}$ a target system operator.
In this representation, the reduced density matrix reads
\begin{equation}
  \bra{0} \chi^{\rm rec} = \int_{-\infty}^\infty dx \, \frac{e^{-x^2/2}}{\pi^{1/4}} \chi_x.
  \label{eq:NumAccuracy_rhoSinx}
\end{equation}
The ladder operators $a$ and $a^\dagger$ read in the position basis as $a \to (x + \partial_x) / \sqrt{2}$ and $a^\dagger \to (x - \partial_x) / \sqrt{2}$ with $\partial_x$ the derivative operator with respect to $x$.
Therefore, the evolution equation of $\chi_x$ is given by $(\partial/\partial t) \chi_x = \bra{x} \bar{\mathcal{M}}(\chi^{\rm rec})$ with
\begin{equation}
\begin{gathered}
  \bra{x} \bar{\mathcal{M}}(\chi^{\rm rec}) = \mathcal{L}_S (\chi_x) + \frac{\kappa}{4} ( \partial_x^2 - x^2 + 1 ) \chi_x \\
  + \sqrt{2} u (x - \partial_x ) \chi_x - i \sqrt{2} x [V_S, \chi_x],
\end{gathered}
  \label{eq:NumAccuracy_xM}
\end{equation}
For initial condition of $\chi_x$,
the vacuum initial state $\chi^{\rm rec}(t=0) = \rho_S(0) \otimes \ket{0}$ corresponds to
\begin{equation*}
  \chi_x = \frac{e^{-x^2/2}}{\pi^{1/4}} \rho_S(0).
\end{equation*}

In numerical simulations, the position space needs to be confined in a finite-size box $x_{\rm min} \leq x < x_{\rm max}$ and be discretized with the grid size $\Delta x$ as $x_m = x_{\rm min} + m \Delta x \ (m = 0,1,\dots,N_{x}-1)$ where $N_{x} =(x_{\rm max} - x_{\rm min})/\Delta x$ is the dimension of the oscillator system space in simulations and corresponds to $N_{\rm tr}$ in the Fock basis.
The evolution equation is integrated by directly computing $\chi^{\rm rec}(t) = \exp(\bar{\mathcal{M}}t) (\chi^{\rm rec} (t=0))$ as in Fig.$\,$\ref{fig:NumAccuracy_errors_ugdep}.
For the numerical differentiations in Eq.$\,$(\ref{eq:NumAccuracy_xM}), we use the 13-point formulas, namely
$\partial_x \chi_x = \sum_{k=1}^6 c_{1,k} (\chi_{x + k \Delta x} - \chi_{x- k \Delta x})$ and
$\partial_x^2 \chi_x = \sum_{k=1}^6 c_{2,k} (\chi_{x + k \Delta x} + \chi_{x- k \Delta x} - 2 \chi_x)$
where the coefficients $\{ c_{l,k} \}$ are determined from the Taylor series of $\{ \chi_{x+ k \Delta t} \}_{k = -6,-5,\dots,5,6}$.
The reduced density matrix is computed as $\rho_S^{\rm comp} = \sum_{m=0}^{N_x - 1} \Delta x (\exp(-x_m^2/2)/\pi^{1/4}) \chi_{x_m}$ (see Eq.$\,$(\ref{eq:NumAccuracy_rhoSinx})).
We confirmed that the numerical error remains the same when different integration methods, such as the compound Simpson's rule, are used in the computation of the reduced density matrix.

\begin{figure}[t]
  \includegraphics[keepaspectratio, scale=0.5]{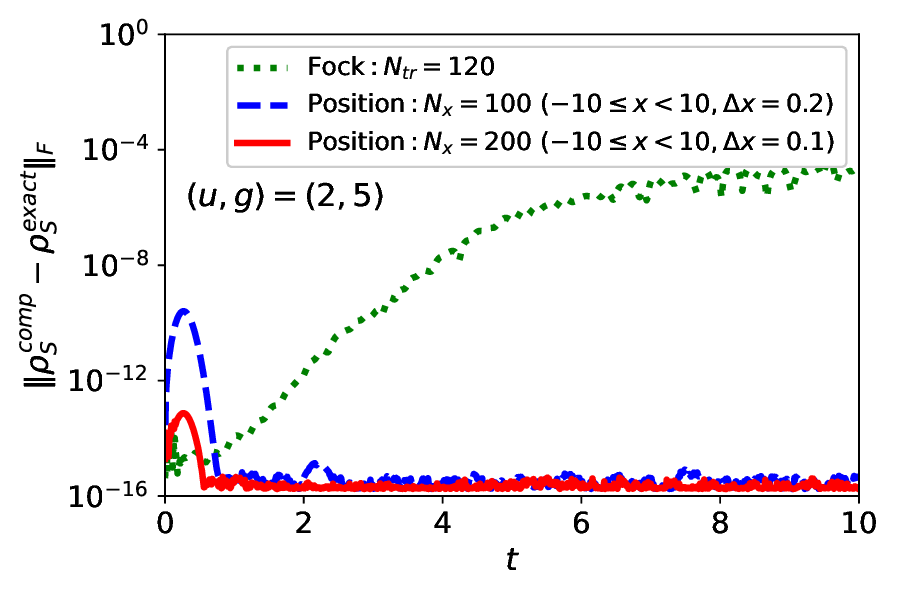}
  \caption{
  Comparison of numerical errors: the position basis (dashed blue and solid red lines) versus the Fock basis (the dotted green line). The time $t$ is shown in the unit of $\kappa = 1$.
  }
  \label{fig:errors_position_A}
\end{figure}

For the parameter values $(u,g) = (2,5)$, the resulting numerical errors are shown in Fig.$\,$\ref{fig:errors_position_A}.
It shows that the numerical errors in the long-time domain are smaller than the plateaued results in the Fock basis.
By performing numerical tests for various other initial states, it turned out that the result behaves as Fig.$\,$\ref{fig:errors_position_A} whenever ${\rm tr}_S(\rho_S(0) \sigma_x) = 0$ or, equivalently, whenever the asymptotic state is given by $I_2/2$.
In such cases, the errors in the long-time domain are reduced to the order of $10^{-16}$.
While the error grows in the short time domain, better accuracy can be achieved by decreasing $\Delta x$.
The other cases in which ${\rm tr}_S(\rho_S(0) \sigma_x) \ne 0$ are explored in Appendix \ref{app:plateau_x}.
In these cases, we could also reduce the numerical errors below the plateau value.

In the Fock basis, we mentioned in Sec.$\,$\ref{subsec:NumAccuracy_plateau} that $\rho_S^{\rm comp}$ diverges with larger values of $u$.
This could not be resolved by increasing the truncation level $N_{\rm tr}$.
In the position basis, on the other hand, we were able to obtain stable results with enlarged position space and small $\Delta x$.
For instance, when $(u,g) = (4,5)$, the errors at $t = 10$ can be reduced to the order of $10^{-10}$ with $N_x = 2000 \ (-20 \leq x < 20, dx = 0.02)$, while the result with the Fock basis blows up even with $N_{\rm tr} = 2000$.

\section{Concluding remarks}
\label{sec:conc}

For the Liouvillian given by Eq.$\,$(\ref{eq:method_Ltot}), we introduced the moment expansion method, which is based on moments of the quadrature operators (see Eqs.$\,$(\ref{eq:method_chirho}) and (\ref{eq:method_LowMoments})).
This new method allows us to reduce the cost in computing the reduced density matrix and the low-order moments of the quadrature operators, including the correlation function, compared with the conventional method.
The reduction of the numerical cost is expected because the degrees of freedom associated with the momentum quadrature are decoupled to those of the position quadrature (see the remark (b) in Sec.$\,$\ref{sec:method}).
Numerical tests were performed to discuss the numerical accuracy of the moment expansion method.
Detailed studies in Sec.$\,$\ref{sec:NumAccuracy} revealed that the accuracy in the long-time domain behaves differently depending on the parameter $u$ (real part of the drive amplitude $\epsilon$).
When $u = 0$, on the one hand, the errors can be reduced to the order of machine precision with a sufficiently large truncation level.
A master equation for an optomechanical setting belongs to this class when the linearization approximation is made on the cavity mode, as shown in Sec.$\,$\ref{sec:OptMech}.
The computation of the correlation functions in this system showed that the moment expansion method could reproduce the results of the conventional method at a less computational cost.
When $u \ne 0$, on the other hand, simulations with the Fock basis exhibit a plateau in the accuracy in the long-time domain.
We found that this plateau problem can be mitigated using the position basis.

As a future perspective, we point out the possibility of applying the moment expansion method to a resonantly driven cavity under the continuous homodyne detection of the momentum quadrature.
In an optomechanical setting, such a situation can be used to generate a conditioned squeezed state in the mechanical oscillator \cite{CFJ08}.
In this case, the stochastic term is coupled only to the degree of freedom associated with the momentum quadrature (right index of $\chi_{m,n}$).
The required truncation level for the right index then might not be as large as that for the left index, and the moment expansion method might still provide an efficient means of simulations.


\begin{acknowledgments}
  I thank Claude Le Bris for his valuable input throughout the research process and for his careful review of the paper.
  I am also grateful to Pierre Rouchon, Lev-Arcady Sellem, and Angela Riva for useful comments during the preparation of this paper.
  This work has been supported by the Engineering for Quantum Information Processors (EQIP) Inria challenge project.
\end{acknowledgments}


\appendix
\section{Transformation to moment expansion representations}
\label{app:RhoXiChi}

We proposed in this article to track the evolution of the operator $\chi$, instead of the density matrix $\rho$.
The operator $\chi$ was introduced in Eq.$\,$(\ref{eq:method_chi}) with its elements $\chi_{m,n}$ defined by Eq.$\,$(\ref{eq:method_chirho}).
The authors of \cite{Link22}, on the other hand, considered elements proportional to ${\rm tr}_{\rm osc} (a^m \rho (a^\dagger)^n )$.
To align with the pre-factor of $\chi_{m,n}$, we introduce the elements as
\begin{equation*}
  \xi_{m,n} = \frac{{\rm tr}_{\rm osc}(a^m \rho (a^\dagger)^n )}{\sqrt{m! \, n!}} \ \ \ (m,n = 0,1,2,\dots),
\end{equation*}
which are the operators in the target system space, and the operator
\begin{equation*}
  \xi = \sum_{m,n = 0}^{\infty} \xi_{m,n} \otimes \ketbra{m}{n},
\end{equation*}
in the total space.
We find $\xi_{m,n}^\dagger = \xi_{n,m}$ and $\chi_{m,n}^\dagger = (-1)^n \chi_{m,n}$.
Let $\mathcal{T}_\eta \ (\eta = \xi \ {\rm or} \ \chi)$ be the map sending $\rho$ to $\eta$, $\eta = \mathcal{T}_\eta (\rho)$.
In this appendix, we reveal several properties of these transformations.

For $\eta = \xi \ {\rm and} \ \chi$, $\mathcal{T}_\eta$ are linear maps.
In addition, $\mathcal{T}_\eta$ apply identically to the target system space.
Thus, with $A_S$ and $B_S$ any operators acting only on the target system space, we have $\mathcal{T}_\eta (A_S \rho B_S) = A_S \eta B_S$.
Furthermore, for any operator $A$,
\begin{equation}
  {\rm tr}_{\rm osc} (A) = \braket{0|\mathcal{T}_\eta (A) |0}.
  \label{eq:RhoXiChi_0T0}
\end{equation}

\subsubsection{Transformation of a Liouvillian}

We first transform a Liouvillian by $\mathcal{T}_\eta$.
For generality, let us consider the following Liouvillian, which is more general than Eq.$\,$(\ref{eq:method_Ltot}),
\begin{equation*}
  \begin{gathered}
    \mathfrak{L}(\rho) = \mathcal{L}_S (\rho) \\
    - i \Delta [a^\dagger a, \rho] + [\epsilon a^\dagger - \epsilon^* a, \rho] + \kappa \mathcal{D}[a] (\rho) \\
    - i [ L_S \, a^\dagger + L_S^\dagger \, a, \rho], \\
  \end{gathered}
\end{equation*}
where $\Delta$ is the oscillator detuning and $L_S$ is a target system operator.
By assuming $\Delta = 0$ and $L_S = V_S$ (Hermitian), we recover Eq.$\,$(\ref{eq:method_Ltot}).
To have a more compact form, we introduce $F_S = L_S + i \epsilon I_S$.
The Liouvillian $\mathfrak{L}$ then reads
\begin{equation*}
  \begin{gathered}
    \mathfrak{L}(\rho) = \mathcal{L}_S (\rho) \\
    - i \Delta [a^\dagger a, \rho] + \kappa \mathcal{D}[a] (\rho)
    - i [ F_S \, a^\dagger + F_S^\dagger \, a, \rho].
  \end{gathered}
\end{equation*}

To find how $\mathfrak{L}$ is transformed by $\mathcal{T}_\eta$, the following relations are convenient.
We have for $\eta = \xi$,
\begin{equation*}
  \begin{gathered}
    \mathcal{T}_\xi (a \rho) = a \xi, \ \
    \mathcal{T}_\xi (a^\dagger \! \rho) = \{ a^\dagger, \xi \}, \\
    \mathcal{T}_\xi (\rho a^\dagger) = \xi a^\dagger, \ \
    \mathcal{T}_\xi (\rho a) = \{ a, \xi \},
  \end{gathered}
\end{equation*}
and for $\eta = \chi$,
\begin{equation}
  \begin{gathered}
    \mathcal{T}_\chi(a \rho) = \frac{1}{2}\Upsilon_+(\chi), \ \
    \mathcal{T}_\chi(a^\dagger \! \rho) = \Big( \frac{1}{2} \Upsilon_- + \Upsilon_+^* \Big) (\chi), \\
    \mathcal{T}_\chi(\rho a^\dagger) = \frac{1}{2}\Upsilon_-(\chi), \ \
    \mathcal{T}_\chi(\rho a) = \Big( \frac{1}{2} \Upsilon_+ + \Upsilon_-^* \Big) (\chi),
  \end{gathered}
  \label{eq:RhoXiChi_Ta}
\end{equation}
with $\Upsilon_\pm^*$ defined by $\Upsilon_\pm^*(\rho) = a^\dagger \rho \pm \rho a$.
For instance, we can show from these relations that the dissipator $\mathcal{D}[a]$ is diagonalized in the Fock basis after the transformations,
\begin{equation}
  \mathcal{T}_\eta (\mathcal{D}[a] (\rho)) = - \frac{1}{2} \{ a^\dagger a, \eta \} \ \ \ (\eta = \chi \ {\rm and} \ \xi).
  \label{eq:RhoXiChi_DiagDissipator}
\end{equation}
The Liouvillian $\mathfrak{L}$ can be transformed as
\begin{equation*}
  \mathfrak{M}_\eta(\eta) = \mathcal{T}_\eta(\mathfrak{L}(\rho)).
\end{equation*}
For $\eta = \xi$, we have
\begin{equation*}
  \begin{gathered}
    \mathfrak{M}_\xi (\xi) = \mathcal{L}_S(\xi) \\
    - i \Delta [a^\dagger a, \xi] - \frac{\kappa}{2} \{ a^\dagger a, \chi \} \\
    - i [ F_S \, a^\dagger + F_S^\dagger \, a, \xi] - i (F_S \xi a^\dagger - a \xi F_S^\dagger), \\
  \end{gathered}
\end{equation*}
and its $(m,n)$ element reads
\begin{equation}
  \begin{gathered}
    \braket{m|\mathfrak{M}_\xi (\xi)|n} = \mathcal{L}_S(\xi_{m,n}) \\
    - i \Delta (m-n) \xi_{m,n} - \frac{\kappa}{2}(m+n) \xi_{m,n} \\
    - i ( \sqrt{m} \, F_S \xi_{m-1,n} - \sqrt{n} \, \xi_{m,n-1} F_S^\dagger) \\
    - i ( \sqrt{m+1} \, [F_S^\dagger, \xi_{m+1,n}] + \sqrt{n+1} \, [F_S, \xi_{m,n+1}] ).
  \end{gathered}
  \label{eq:RhoXiChi_Mximn}
\end{equation}
For $\eta = \chi$, on the other hand, we have
\begin{equation}
  \begin{gathered}
    \mathfrak{M}_\chi(\chi) = \mathcal{L}_S(\chi) -i \Delta (a \chi a + a^\dagger \chi a^\dagger) - \frac{\kappa}{2} \{ a^\dagger a, \chi \} \\
    - i a^\dagger (F_S \chi - \chi F_S^\dagger) - i (F_S \chi + \chi F_S^\dagger) a \\
    - \frac{i}{2} [F_S + F_S^\dagger,a \chi]
     + \frac{i}{2} [F_S - F_S^\dagger, \chi a^\dagger],
  \end{gathered}
  \label{eq:RhoXiChi_Mchi}
\end{equation}
and its $(m,n)$ element reads
\begin{equation}
  \begin{gathered}
    \braket{m|\mathfrak{M}_\chi(\chi)|n} = \mathcal{L}_S(\chi_{m,n}) - \frac{\kappa}{2}(m+n) \chi_{m,n} \\
    - i \sqrt{m} (F_S \chi_{m-1,n} - \chi_{m-1,n} F_S^\dagger) \\
    - i \sqrt{n} (F_S \chi_{m,n-1} + \chi_{m,n-1} F_S^\dagger) \\
    - \frac{i}{2} \sqrt{m+1} [F_S + F_S^\dagger, \chi_{m+1,n}] \\
    - i \Delta (\sqrt{n(m+1)} \chi_{m+1,n-1} + \sqrt{m(n+1)} \chi_{m-1,n+1}) \\
    + \frac{i}{2} \sqrt{n+1} [F_S - F_S^\dagger, \chi_{m,n+1}].
  \end{gathered}
  \label{eq:RhoXiChi_Mchimn}
\end{equation}

Let us examine the dependence on the elements with higher indices than $(m,n)$ as in the remark (b) underneath Eq.$\,$(\ref{eq:method_dchimn}).
In Eq.$\,$(\ref{eq:RhoXiChi_Mximn}), the elements having higher indices than $\xi_{m,n}$ are $\xi_{m+1,n}$ and $\xi_{m,n+1}$.
Those elements come with the $F_S$ operator and thus result from the terms involving $\epsilon$ and the interaction term.
The terms with $\epsilon$ do not contribute because of the commutators, and we have $[L_S^\dagger, \xi_{m+1,n}]$ and $[L_S, \xi_{m,n+1}]$.
Thus, $\xi_{m,n}$ is inevitably coupled to both $\xi_{m+1,n}$ and $\xi_{m,n+1}$.
In Eq.$\,$(\ref{eq:RhoXiChi_Mchimn}), on the other hand, the elements having higher indices than $\chi_{m,n}$ are $\chi_{m+1,n-1}$, $\chi_{m-1,n+1}$, $\chi_{m+1,n}$, and $\chi_{m,n+1}$.
The first two elements $\chi_{m+1,n-1}$ and $\chi_{m-1,n+1}$ result from the term involving $\Delta$, and they are eliminated when we assume that $\Delta = 0$.
The latter two elements $\chi_{m+1,n}$ and $\chi_{m,n+1}$ result from the terms involving $\epsilon$ and the interaction term.
The commutators remove the contribution of the terms with $\epsilon$, and we have $[L_S + L_S^\dagger, \chi_{m+1,n}]$ and $[L_S - L_S^\dagger, \chi_{m,n+1}]$.
In this case, thus, we can eliminate the term with $\chi_{m,n+1}$ by assuming $L_S = V_S$ (Hermitian).

Eq.$\,$(\ref{eq:RhoXiChi_Mximn}) is similar to the hierarchical equation of motion Eq.$\,$(13) in \cite{Link22}.
Slight differences originate from the fact that the authors defined their elements as $\bar{\xi}_{m,n} = {\rm tr}_{\rm osc} ((iga)^m \rho (-iga^\dagger)^n )$, where $g$ is the coupling constant (in \cite{Link22}, the interaction Hamiltonian is of the form $g (L \, a^\dagger + L^\dagger \, a)$ with a dimensionless operator $L$).
Using the relation $\bar{\xi}_{m,n} = (ig)^m (-ig)^n \sqrt{m! n!} \, \xi_{m,n}$, we can derive Eq.$\,$(13) in \cite{Link22} from Eq.$\,$(\ref{eq:RhoXiChi_Mximn}).

\subsubsection{Inverse transformations}

We next show the existence of the inverse of $\mathcal{T}_\eta \ (\eta = \xi \ {\rm and} \ \chi)$.
For this, it is sufficient to show that the density matrix $\rho$ can be constructed from $\{ \eta_{m,n} \}$.

We begin with
\begin{equation}
  \rho_{m,n} = \braket{m|\rho|n} = \frac{ {\rm tr}_{\rm osc}( \ketbra{0}{0} a^m \rho (a^\dagger)^n )}{\sqrt{m! n!}}.
  \label{eq:RhoXiChi_RhoTr}
\end{equation}
One can directly show $\braket{m| \sum_{l = 0}^\infty ((-1)^l / l!) (a^\dagger)^l a^l |n} = \delta_{m,0} \delta_{n,0}$, which leads to
$\ketbra{0}{0} = \sum_{l = 0}^\infty ((-1)^l / l!) [a^\dagger]^l a^l$.
Inserting this relation into Eq.$\,$(\ref{eq:RhoXiChi_RhoTr}) yields
\begin{equation}
  \rho_{m,n} = \sum_{l=0}^\infty (-1)^l \frac{ {\rm tr}_{\rm osc}( a^{m+l} \rho (a^\dagger)^{n+l} ) }{l!\sqrt{m! n!}},
  \label{eq:RhoXiChi_RhoTr1}
\end{equation}
or
\begin{equation*}
  \rho_{m,n} = \sum_{l=0}^\infty \frac{(-1)^l}{l!} \sqrt{\frac{(m+l)!(n+l)!}{m!n!}} \xi_{m+l,n+l}.
\end{equation*}
Thus, with Eq.$\,$(\ref{eq:method_rho}), $\rho$ can be constructed from $\{ \xi_{m,n} \}$.

From the definition of $\Upsilon_{\rm \pm}$, we find $a \rho = (\Upsilon_+ + \Upsilon_-)(\rho)/2$, $\rho a^\dagger = (\Upsilon_+ - \Upsilon_-)(\rho)/2$.
Inserting these relations into Eq.$\,$(\ref{eq:RhoXiChi_RhoTr1}) yields
\begin{equation*}
  \begin{gathered}
    \rho_{m,n} = \sum_{l=0}^\infty \frac{(-1)^l}{l!\sqrt{m! n!}} \\
    \times {\rm tr}_{\rm osc} \left(  \Big[ \frac{\Upsilon_+ + \Upsilon_-}{2} \Big]^{m+l} \Big[ \frac{\Upsilon_+ - \Upsilon_-}{2} \Big]^{n+l} (\rho) \right) \\
    = \sum_{l=0}^\infty \sum_{p=0}^{m+l} \sum_{q=0}^{n+l} \frac{(-1)^{n-q}}{l!\sqrt{m! n!}  \,  2^{2l+m+n} } {m+l \choose p} {n+l \choose q} \\
    \times {\rm tr}_{\rm osc} ( \Upsilon_+^{p+q} \Upsilon_-^{2l+m+n-p-q} (\rho) ).
  \end{gathered}
\end{equation*}
with the binomial coefficient ${a \choose b} = a!/(b! (a-b)!)$.
This equation leads to
\begin{equation}
  \begin{gathered}
    \rho_{m,n} = \sum_{l=0}^\infty \sum_{p=0}^{m+l} \sum_{q=0}^{n+l} \frac{(-1)^{n-q}}{l! \, 2^{2l+m+n} } \\
    \times \sqrt{ \frac{(p+q)! (2l+m+n-p-q)!}{m!n!} } \\
    \times {m+l \choose p} {n+l \choose q} \, \chi_{p+q,2l+m+n-p-q},
  \end{gathered}
  \label{eq:RhoXiChi_RhoFromChi}
\end{equation}
which shows that $\rho$ can be constructed from $\{ \chi_{m,n} \}$.

\subsubsection{Correlation function}

We next consider how to calculate the correlation function in the new representation.
Here we focus only on $\chi$. The following formulation can also be performed in the representation $\xi$.
Suppose that the application of an operator $O$ from the left is transformed to an operation $\mathcal{O}$ by $\mathcal{T}_\chi$, namely, $\mathcal{T}_\chi(OA) = \mathcal{O}(\mathcal{T}_\chi(A))$ for any operator $A$.
With the aid of Eq.$\,$(\ref{eq:RhoXiChi_0T0}), the correlation function Eq.$\,$(\ref{eq:OptMech_c_o}) can then be calculated with $\chi$ as
\begin{equation*}
  \begin{gathered}
    C_O(t) = {\rm tr} (O e^{\mathfrak{L} t}( O \rho_{ss})) - [{\rm tr} (O \rho_{ss} )]^2 \\
    = {\rm tr}_S \braket{0| \mathcal{T}_\chi(O e^{\mathfrak{L} t} ( O \rho_{ss} ) ) |0} - [{\rm tr}_S \braket{0| \mathcal{T}_\chi(O \rho_{ss}) |0}]^2 \\
    = {\rm tr}_S \braket{0| \mathcal{O}( e^{\mathfrak{M}_\chi t} ( \mathcal{O} (\chi_{ss})))|0} - [{\rm tr}_S \braket{0| \mathcal{O} (\chi_{ss}) |0}]^2,
  \end{gathered}
\end{equation*}
where $\chi_{ss}$ is the steady state in the $\chi$ representation, $\chi_{ss} = \mathcal{T}_\chi(\rho_{ss}) = \lim_{t \to \infty} e^{\mathfrak{M}_\chi t} (\chi)$.

For instance, when $O$ acts only on the target system space, $\mathcal{O}(A) = OA$ for any $A$ and thus
\begin{equation}
  \begin{gathered}
    C_O(t) = {\rm tr}_S (O \braket{0| e^{\mathfrak{M}_\chi t} ( O \chi_{ss} )|0}) \\
    - [{\rm tr}_S (O \braket{0|\chi_{ss}|0} )]^2.
  \end{gathered}
  \label{eq:RhoXiChi_autocorr_sys}
\end{equation}
When $O$ is the momentum quadrature operator of the oscillator mode, $P_{\rm osc}= i(a^\dagger-a)/\sqrt{2}$, we can show that $\mathcal{O}(A) = (i/\sqrt{2}) a^\dagger A - A P_{\rm osc}$ from Eqs.$\,$(\ref{eq:RhoXiChi_Ta}). Therefore,
\begin{equation}
  \begin{gathered}
    C_{P_{\rm osc}} (t) = \frac{i}{\sqrt{2}} {\rm tr}_S \braket{0| e^{\mathfrak{M}_\chi t} ( \chi_{ss} P_{\rm osc}- \frac{i}{\sqrt{2}} a^\dagger \chi_{ss} ) |1} \\
    + \frac{1}{2} [{\rm tr}_S \braket{0|\chi_{ss}|1} ]^2.
  \end{gathered}
  \label{eq:RhoXiChi_autocorr_p_osc}
\end{equation}


\section{Determination of the truncation levels in Sec.$\,$\ref{sec:OptMech}}
\label{app:det.trunc}

The truncation levels $N_{\rm osc}$ and $N_{\rm mec}$ in Sec.$\,$\ref{sec:OptMech} are determined by checking convergence of the steady state computation as follows.
With $C_O^{N_{\rm osc}, N_{\rm mec}}(t=0)$ being the correlation function at $t = 0$ computed with the truncation levels $N_{\rm osc}$ and $N_{\rm mec}$, we estimate the errors using
\begin{equation*}
  \begin{gathered}
    \Delta_O^{(1)}(N_{\rm osc},N_{\rm mec}) = \\
    \frac{|C_O^{N_{\rm osc}, N_{\rm mec}}(t=0) - C_O^{N_{\rm osc}+3, N_{\rm mec}}(t=0)|}{|C_O^{N_{\rm osc}, N_{\rm mec}}(t=0)|},
  \end{gathered}
\end{equation*}
and
\begin{equation*}
  \begin{gathered}
    \Delta_O^{(2)}(N_{\rm osc},N_{\rm mec}) = \\
    \frac{|C_O^{N_{\rm osc}, N_{\rm mec}}(t=0) - C_O^{N_{\rm osc}, N_{\rm mec}+3}(t=0)|}{|C_O^{N_{\rm osc}, N_{\rm mec}}(t=0)|}.
  \end{gathered}
\end{equation*}
Starting with $N_{\rm osc} = N_{\rm mec} = 3$, we increase them by 3 until $\Delta_O^{(i)} < 10^{-4}$ is reached for $i = 1,2$ and $O = X_{\rm mec}, P_{\rm osc}$.
We then use the final values of $N_{\rm osc}$ and $N_{\rm mec}$ in our simulations.

\section{Exact solution to the eigenvalue problem for $\bar{\mathcal{M}}$ and $\mathcal{L}$ when $\mathcal{L}_S = 0$}
\label{app:LDiag}

For finite-dimensional target systems, the dimension of which is denoted by $d_S$, the exact spectrum and eigenoperators for the generators $\mathcal{L}$ Eq.$\,$(\ref{eq:method_Ltot}) and $\bar{\mathcal{M}}$ Eq.$\,$(\ref{eq:method_M}) can be obtained analytically when $\mathcal{L}_S = 0$.
From those solutions, we can calculate the exact solutions to the evolution equations Eqs.$\,$(\ref{eq:method_Leq}) and (\ref{eq:method_barchi}) for any initial state in principle.
In this appendix, we present diagonalization procedures for $\bar{\mathcal{M}}$ and $\mathcal{L}$.

\subsection{Diagonalization of $\bar{\mathcal{M}}$}
\label{app:LDiag_M}

Throughout this appendix, we employ the Hilbert-Schmidt inner product.
For rectangular matrices of the form Eq.$\,$(\ref{eq:method_barchi}), it reads $(\chi^{\rm rec}_1, \chi^{\rm rec}_2) = {\rm tr}_S ((\chi^{\rm rec}_1)^\dagger \chi^{\rm rec}_2)$.
Given a superoperator $\mathcal{B}$ on the rectangular matrices, its adjoint $\mathcal{B}^{\rm ad}$ is introduced as the superoperator that satisfies
$(\chi^{\rm rec}_1, \mathcal{B}(\chi^{\rm rec}_2)) = {\rm tr}_S ( (\chi^{\rm rec}_1)^\dagger \mathcal{B}(\chi^{\rm rec}_2) ) = (\mathcal{B}^{\rm ad} (\chi^{\rm rec}_1), \chi^{\rm rec}_2) = {\rm tr}_S( (\mathcal{B}^{\rm ad}(\chi^{\rm rec}_1))^\dagger \chi^{\rm rec}_2)$
for $\chi^{\rm rec}_1$ and $\chi^{\rm rec}_2$ any rectangular matrices of the form Eq.$\,$(\ref{eq:method_barchi}).

We consider the eigenvalue problem for $\bar{\mathcal{M}}$ when $\mathcal{L}_S = 0$;
\begin{equation*}
  (\bar{\mathcal{M}} - \lambda) (_R \theta^{\rm rec}) = 0, \ \ \ (\bar{\mathcal{M}}^{\rm ad} - \lambda^*) (_L \theta^{\rm rec}) = 0.
\end{equation*}
When $\mathcal{L}_S = 0$, the operator $V_S$ is the only target system operator appeared in $\bar{\mathcal{M}}$.
Suppose that the eigenvalue problem for $V_S$ is solved as
\begin{equation}
  V_S \ket{v_i} = g_i \ket{v_i} \ \ \ \ \ \ \ (i = 1,\dots,d_S).
  \label{eq:LDiag_VS}
\end{equation}
Since $V_S$ is Hermitian, the eigenvalues $\{ g_i \}$ are real and the eigenvectors are orthonormal, $\braket{v_i|v_j} = \delta_{i,j}$ for $i,j = 1,\dots,d_S$.
With $\{ \ket{v_i} \}$, we have for any oscillator state $\ket{\psi_{\rm osc}}$,
\begin{equation}
  \bar{\mathcal{M}} (\ketbra{v_i}{v_j} \otimes \ket{\psi_{\rm osc}} ) = \ketbra{v_i}{v_j} \otimes M_{i,j}^u \ket{\psi_{\rm osc}},
  \label{eq:LDian_Mintro}
\end{equation}
where $M_{i,j}^u$ is defined by
\begin{equation*}
  M_{i,j}^u = - \frac{\kappa}{2} a^\dagger a + 2u a^\dagger - i g_{i,j} (a+a^\dagger).
\end{equation*}
This operator can be diagonalized as
\begin{equation}
  W_{i,j}^{-1}  M_{i,j}^u W_{i,j} = - \frac{\kappa}{2} (a^\dagger a + z_{i,j}),
  \label{eq:LDiag_Mphoton_diag}
\end{equation}
where $z_{i,j} = 4 g_{i,j} (g_{i,j} + 2iu)/\kappa^2$ and the invertible operator $W_{i,j}$ is defined by
\begin{equation*}
  W_{i,j} = D \Big( \frac{2ig_{i,j}}{\kappa} \Big) \exp \Big( \frac{4(u-ig_{i,j})}{\kappa} a^\dagger \Big).
\end{equation*}
with the displacement operator $D(\alpha) = \exp(\alpha a^\dagger - \alpha^* a)$.
Therefore, for $i,j = 1,\dots,d_S$ and $m = 0,1,2,\dots$, the eigenvalues, the right eigenoperators, and the left eigenoperators are respectively given by
\begin{equation*}
  \lambda_{i,j,m} = - \frac{\kappa}{2} (m + z_{i,j}),
\end{equation*}
\begin{equation*}
  _R \theta_{i,j,m}^{\rm rec} = \ketbra{v_i}{v_j} \otimes W_{i,j} \ket{m}, \\
\end{equation*}
and
\begin{equation*}
  _L \theta_{i,j,m}^{\rm rec} = \ketbra{v_i}{v_j} \otimes (W_{i,j}^{-1})^\dagger \ket{m}.
\end{equation*}
Using the orthonormality and the completeness of the Fock basis, we can show the same properties for the set $\{ _R \theta_{i,j,m}^{\rm rec}, \ _L \theta_{i,j,m}^{\rm rec} \}$.
Note that ${\rm Re}(\lambda_{i,j,m}) = - (\kappa/2) (m + 4 g_{i,j}^2/\kappa^2) \leq 0$, indicating the stability of the dynamics.

The formal solution to the evolution equation Eq.$\,$(\ref{eq:method_masterchi}) reads $\chi^{\rm rec}(t) = \exp(\bar{\mathcal{M}} t) (\chi^{\rm rec}(t = 0))$.
Thus, using the spectral decomposition of $\exp(\bar{\mathcal{M}} t)$, we obtain
\begin{equation*}
  \chi^{\rm rec}(t) = \sum_{i,j = 1}^{d_S} \sum_{m = 0}^\infty e^{\lambda_{i,j,m} t} {\rm tr}_S( (_L \theta^{\rm rec}_{i,j,m})^\dagger \chi^{\rm rec}(t=0) ) _R \theta^{\rm rec}_{i,j,m}.
\end{equation*}
The exact evolution of the reduced density matrix can be calculated by taking the zeroth element.
If the oscillator is initially in the vacuum state as $\chi^{\rm rec}(t=0) = \rho_S(0) \otimes \ket{0}$, then we find
\begin{equation}
  \begin{gathered}
    \rho_S^{\rm exact} (t) = \\
    \sum_{i,j = 1}^{d_S} e^{z_{i,j} (1 - \kappa t /2 - e^{- \kappa t /2})} \ \ketbra{v_i}{v_i} \rho_S (0) \ketbra{v_j}{v_j}.
  \end{gathered}
  \label{eq:LDiag_ExactRhoS}
\end{equation}

From here on, we assume that all the eigenvalues of $V_S$ are nondegenerate.
Note that this is true for the two-level example in Sec.$\,$\ref{sec:NumAccuracy},
Then, we see that $\lambda_{i,j,m} = 0$ if and only if $i = j$ and $m = 0$.
Therefore, the generator on the oscillator system space responsible for the long-time domain (after decay of all the modes with ${\rm Re} (\lambda_{i,j,m}) < 0$) is $M_{i,i}^u = - (\kappa/2) a^\dagger a + 2 u a^\dagger$, which is independent of the target system state index $i$.

\subsection{Diagonalization of $\mathcal{L}$}
\label{app:LDiag_L}

For total state operators, the Hilbert-Schmidt inner product reads $(\rho_1, \rho_2) = {\rm tr}(\rho_1^\dagger \rho_2)$.
Given a superoperator $\mathcal{A}$, its adjoint $\mathcal{A}^{\rm ad}$ satisfies
$(\rho_1, \mathcal{A}(\rho_2)) = {\rm tr}(\rho_1^\dagger \mathcal{A}(\rho_2)) = (\mathcal{A}^{\rm ad} (\rho_1), \rho_2) = {\rm tr}((\mathcal{A}^{\rm ad}(\rho_1))^\dagger \rho_2))$
for $\rho_1$ and $\rho_2$ any operators in the total space

It is instructive to start with the eigenvalue problem for $\mathcal{D}[a]$.
It can be solved exactly using, for instance, the ladder superoperator technique \cite{Prosen10,GP17,BZ21}, which proceeds as follows.
We introduce the superoperators
\begin{equation*}
  \begin{gathered}
    \mathcal{A}_0 (\rho ) = a \rho, \ \mathcal{A}_1 (\rho ) = \rho a^\dagger, \\
    \mathcal{A}_0' (\rho ) = [a^\dagger, \rho], \ \mathcal{A}_1' (\rho ) = [\rho, a],
  \end{gathered}
\end{equation*}
which satisfy
\begin{equation}
  \begin{gathered}
    [\mathcal{A}_\mu, \mathcal{A}_\nu'] = \delta_{\mu,\nu}, \\
    [\mathcal{A}_\mu, \mathcal{A}_\nu] = [\mathcal{A}_\mu', \mathcal{A}_\nu'] = 0, \\
  \end{gathered}
  \label{eq:LDiag_commu}
\end{equation}
for $\mu,\nu = 0,1$. Note that
\begin{equation}
  \mathcal{D}[a] =  - \frac{1}{2} \sum_{\mu = 0}^1 \mathcal{A}_\mu' \mathcal{A}_\mu.
  \label{eq:LDiag_Da}
\end{equation}
While $\mathcal{A}_\mu'$ is not the adjoint of $\mathcal{A}_\mu$,
Eq.$\,$(\ref{eq:LDiag_commu}) resemble the boson commutation relations.
For this reason, these operators are called the ladder superoperators.
Recall that the eigenvectors of $a^\dagger a$, the Fock basis vectors, are constructed using only the commutation relation $[a,a^\dagger] = 1$.
Similarly, we can construct the eigenoperators of $\sum_{\mu = 0}^1 \mathcal{A}_\mu' \mathcal{A}_\mu$, and thus of $\mathcal{D}[a]$ (see Eq.$\,$(\ref{eq:LDiag_Da})), using the commutation relations Eq.$\,$(\ref{eq:LDiag_commu}).
Since $\sum_{\mu = 0}^1 \mathcal{A}_\mu' \mathcal{A}_\mu$ is not self-adjoint, the right and left eigenoperators are different.

To construct the eigenoperators of $\sum_{\mu = 0}^1 \mathcal{A}_\mu' \mathcal{A}_\mu$, we first need vacuum states for the ladder superoperators.
The reader might notice that the Fock vacuum state $\ketbra{0}{0}$ is annihilated by $\mathcal{A}_\mu \ (\mu = 0,1)$,
\begin{equation*}
  \mathcal{A}_\mu (\ketbra{0}{0}) = 0 \ \ \ (\mu = 0,1).
\end{equation*}
Thus, the operators defined by
\begin{equation}
  _R X_{m,n} = \frac{[\mathcal{A}_0']^m}{\sqrt{m!}} \frac{[\mathcal{A}_1']^n}{\sqrt{n!}} (\ketbra{0}{0}) \ \ \ (m,n = 0,1,2,\dots),
  \label{eq:LDiag_XR}
\end{equation}
are the right eigenoperators of $\sum_{\mu = 0}^1 \mathcal{A}_\mu' \mathcal{A}_\mu$;
\begin{equation*}
  \sum_{\mu = 0}^1 \mathcal{A}_\mu' \mathcal{A}_\mu (_R X_{m,n}) = (m + n) _R X_{m,n}.
\end{equation*}
To find the left eigenoperators, we first note
\begin{equation*}
  \mathcal{A}_0'^{\rm ad} (\rho ) = [a, \rho], \ \ \ \mathcal{A}_1'^{\rm ad} (\rho ) = [\rho, a^\dagger].
\end{equation*}
The identity operator $I_{\rm osc}$ is annihilated by these superoperators
\begin{equation*}
  \mathcal{A}_\mu'^{\rm ad} (I_{\rm osc}) = 0 \ \ \ (\mu = 0,1).
\end{equation*}
Thus, the operators defined by
\begin{equation}
  _L X_{m,n} = \frac{[\mathcal{A}_0^{\rm ad}]^m}{\sqrt{m!}} \frac{[\mathcal{A}_1^{\rm ad}]^n}{\sqrt{n!}} (I) \ \ \ (m,n = 0,1,2,\dots),
  \label{eq:LDiag_XL}
\end{equation}
are the left eigenoperators of $\sum_{\mu = 0}^1 \mathcal{A}_\mu' \mathcal{A}_\mu$;
\begin{equation*}
  \Big( \sum_{\mu = 0}^1 \mathcal{A}_\mu' \mathcal{A}_\mu \Big)^{\rm ad} (_L X_{m,n}) = (m + n) _L X_{m,n}.
\end{equation*}
Using the orthonormality $\braket{m|n} = \delta_{m,n}$ and the completeness relation $\sum_{m = 0}^\infty \ketbra{m}{m} = I_{\rm osc}$ of the Fock basis,
we can show the same properties for the set $\{ _R X_{m,n}, \ _L X_{m,n} \}$.

Now consider the eigenvalue problem for $\mathcal{L}$ when $\mathcal{L}_S = 0$;
\begin{equation*}
  (\mathcal{L} - \lambda) (_R \Theta) = 0, \ \ \ (\mathcal{L}^{\rm ad} - \lambda^*) (_L \Theta) = 0.
\end{equation*}
With $\{ \ket{v_i} \}$ in Eq.$\,$(\ref{eq:LDiag_VS}), we have for any oscillator system operator $\rho_{\rm osc}$
\begin{equation*}
  \mathcal{L} (\ketbra{v_i}{v_j} \otimes \rho_{\rm osc}) = \ketbra{v_i}{v_j} \otimes \mathcal{L}_{i,j} (\rho_{\rm osc})
\end{equation*}
where $\mathcal{L}_{i,j}$ is defined by
\begin{equation*}
  \begin{gathered}
    \mathcal{L}_{i,j} (\rho) = [\epsilon a^\dagger - \epsilon^* a, \rho] + \kappa \mathcal{D}[a] (\rho) \\
    - i g_i (a + a^\dagger) \rho + i g_j \rho (a + a^\dagger).
  \end{gathered}
\end{equation*}
Using the ladder superoperators introduced above, we obtain
\begin{equation*}
  \begin{gathered}
    \mathcal{L}_{i,j} = - \frac{\kappa}{2} \sum_{\mu = 0}^1 \mathcal{A}_\mu' \mathcal{A}_\mu \\
    - i g_{i,j} \sum_{\mu = 0}^1 \mathcal{A}_\mu + (\epsilon - i g_i) \mathcal{A}_0' + (\epsilon - i g_j)^* \mathcal{A}_1',
  \end{gathered}
\end{equation*}
with $g_{i,j} = g_i - g_j$.
The terms linear to the ladder superoperators can be eliminated by the similarity transformation as
\begin{equation}
  \mathcal{W}_{i,j}^{-1} \mathcal{L}_{i,j} \mathcal{W}_{i,j} = - \frac{\kappa}{2} \sum_{\mu = 0}^1 \mathcal{A}_\mu' \mathcal{A}_\mu,
  \label{eq:LDiag_W-1LW}
\end{equation}
with the invertible superoperator
\begin{equation*}
  \mathcal{W}_{i,j} = \exp \Big( \frac{2}{\kappa} (i g_{i,j} \sum_{\mu = 0}^1 \mathcal{A}_\mu + (\epsilon - i g_i) \mathcal{A}_0' + (\epsilon - i g_j)^* \mathcal{A}_1') \Big).
\end{equation*}
For instance, the diagonal entries $\mathcal{W}_{i,i}$ is a displacement unitary transformation,
\begin{equation}
  \mathcal{W}_{i,i} (\rho) = D \Big( \frac{2(\epsilon - ig_i)}{\kappa} \Big) \rho D \Big( \frac{2(\epsilon - ig_i)}{\kappa}  \Big)^\dagger.
  \label{eq:LDiag_Wii}
\end{equation}

The eigenvalue problem for $\mathcal{L}$ is solved using Eqs.$\,$(\ref{eq:LDiag_XR}), (\ref{eq:LDiag_XL}), and (\ref{eq:LDiag_W-1LW}).
For $i,j = 1,\dots,d_S$ and $m,n = 0,1,2,\dots$, the eigenvalues, the right eigenoperators, and the left eigenoperators are respectively given by
\begin{equation*}
  \lambda_{i,j,m,n} = - \frac{\kappa}{2} (m + n) - \frac{2g_{i,j}^2}{\kappa} - \frac{4i g_{i,j} {\rm Re}(\epsilon)}{\kappa},
\end{equation*}
\begin{equation*}
  _R \Theta_{i,j,m,n} = \ketbra{v_i}{v_j} \otimes \mathcal{W}_{i,j} (_R X_{m,n}),
\end{equation*}
and
\begin{equation*}
  _L \Theta_{i,j,m,n}^\dagger = \ketbra{v_i}{v_j} \otimes \Big( \mathcal{W}_{i,j}^{-1} \Big)^{\rm ad} (_L X_{m,n}).
\end{equation*}
Using the orthonormality and the completeness of $\{ \ket{v_i} \}$, $\{ _R X_{m,n}, \ _L X_{m,n} \}$, we can show the same properties for the set $\{ _R \Theta_{i,j,m,n}, \ _L \Theta_{i,j,m,n} \}$.

The formal solution to the master equation Eq.$\,$(\ref{eq:method_Leq}) reads $\rho(t) = \exp(\mathcal{L} t) (\rho(t = 0))$.
Thus, using the spectral decomposition of $\exp(\mathcal{L} t)$, we obtain
\begin{equation*}
  \rho(t) = \sum_{i,j = 1}^{d_S} \sum_{m,n = 0}^\infty e^{\lambda_{i,j,m,n} t} {\rm tr}( _L \Theta_{i,j,m,n}^\dagger \rho(t = 0) ) _R \Theta_{i,j,m,n}.
\end{equation*}
The exact evolution of the reduced density matrix can be calculated by taking the partial trace ${\rm tr}_{\rm osc}$.
For the initial vacuum states, we obtained the result identical to Eq.$\,$(\ref{eq:LDiag_ExactRhoS}).

When all the eigenvalues of $V_S$ are simple, $\lambda_{i,j,m,n} = 0$ if and only if $i = j$ and $m = n = 0$.
Thus, the long-time behavior is governed by the generators $\{ \mathcal{L}_{i,i} \}_{i = 1,\dots,d_S}$.

\section{Numerical tests for the plateau problem}
\label{app:plateau}

In Sec.$\,$\ref{subsec:NumAccuracy_plateau}, we found the plateau in numerical accuracy when the parameter $u$ in Eq.$\,$(\ref{eq:method_dchim0}) is nonzero.
This appendix is dedicated to presenting the results of various numerical tests conducted to both examine and address the plateau problem.

\subsection{Dependence on the truncation level $N_{\rm tr}$ and the initial two-level state $\rho_S(0)$}
\label{app:plateau_detail}

\begin{figure}[t]
  \includegraphics[keepaspectratio, scale=0.55]{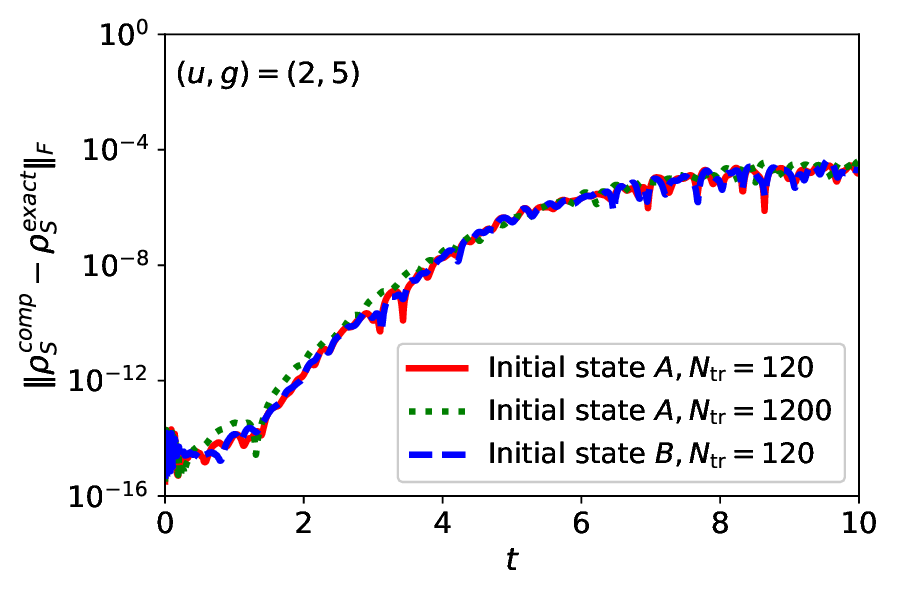}
  \caption{
  Insensitivity of the numerical errors in the long-time domain
  to increasing truncation levels (solid red and dotted green lines)
  and to changing initial two-level states (solid red and dashed blue lines).
 The initial states $A$ and $B$ in the legend are given in Eq.$\,$(\ref{eq:LDiag_Initialtwo-levelAB}). The time $t$ is shown in the unit of $\kappa = 1$.
  }
  \label{fig:NumAccuracy_errors_NtrInitial}
\end{figure}

One might argue that the observed plateau could be resolved with a truncation level much larger than those employed in Figs.$\,$\ref{fig:NumAccuracy_errors_ugdep} $(c)$ and $(d)$.
To test this hypothesis, we perform the simulation with $N_{\rm tr} = 1200$ for the parameter values $(u,g) = (2,5)$ (the same parameter values as Fig.$\,$\ref{fig:NumAccuracy_errors_ugdep} $(d)$ but with 10 times larger $N_{\rm tr}$ than the solid red line in that figure).
The resulting numerical error is shown by the dotted green line in Fig.$\,$\ref{fig:NumAccuracy_errors_NtrInitial}.
The curve closely follows the result with $N_{\rm tr} = 120$ (the solid red line in the figure), thereby indicating that increasing the truncation level does not improve the numerical accuracy.

To see the dependence on the initial two-level state, $\rho_S(0)$, we conduct simulations with the two different initial states,
\begin{align}
  \begin{split}
    {\rm Initial \ state \ A}: \rho_S^{(A)}(0) &= \frac{1}{2} (I_2 - \sigma_z) \ \ {\rm and} \\
    {\rm Initial \ state \ B}: \rho_S^{(B)}(0) &= \frac{1}{2} (I_2 + \sum_{i=x,y,z} \sigma_i/\sqrt{3}).
  \end{split}
  \label{eq:LDiag_Initialtwo-levelAB}
\end{align}
The initial state $A$ corresponds to the one considered in the main text (see Eq.$\,$(\ref{eq:NumAccuracy_Initialtwo-level})).
The dashed blue line in Fig.$\,$\ref{fig:NumAccuracy_errors_NtrInitial} shows the result for the initial state $B$ in Eq.$\,$(\ref{eq:LDiag_Initialtwo-levelAB}).
We see that the overall behavior is similar to the solid red line in the figure showing the result for the initial state $A$.
Other initial two-level states were also examined and found to have the same behavior.
Consequently, the emergence of the plateau is independent of the specific configuration of the initial two-level state.

\subsection{Dependence on the machine precision}
\label{app:plateau_mp}

\begin{figure}[t]
  \includegraphics[keepaspectratio, scale=0.55]{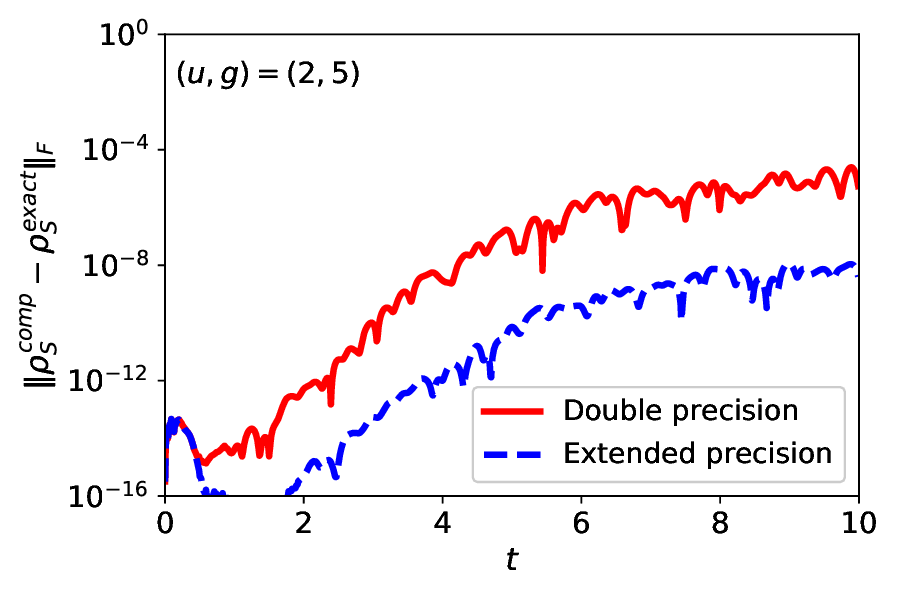}
  \caption{
  Influence of machine precision on the numerical errors.
  The results using floating point numbers in double and extended precision are shown by the solid red and dashed blue lines, respectively.
  The time $t$ is shown in the unit of $\kappa = 1$.
 }
  \label{fig:NumAccuracy_errors_64vs80}
\end{figure}

When using the Fock basis, we have only been able to remedy the plateau problem by increasing machine precision.
Indeed, Fig.$\,$\ref{fig:NumAccuracy_errors_64vs80} compares the results with the double precision floating-point numbers, the 64-bit format which typically gives 16 significant digits, and the extended precision floating-point numbers, the 80-bit format which typically gives 18 significant digits.
When integrating the evolution equation, this time we use the fourth-order Runge-Kutta method with the time step $\Delta t = 10^{-4}$, not the formal solution $\chi^{\rm rec}(t) = \exp(\bar{\mathcal{M}}t) (\chi^{\rm rec} (t=0))$ as in Fig.$\,$\ref{fig:NumAccuracy_errors_ugdep}, because the matrix exponentiation function ${\rm scipy.linalg.expm}$ is not compatible with the extended precision.
Fig.$\,$\ref{fig:NumAccuracy_errors_64vs80} shows the resulting numerical errors.
We first note the similarity between the Runge-Kutta result (solid red line in this figure) and the matrix exponentiation result (solid red line in Fig.$\,$\ref{fig:NumAccuracy_errors_NtrInitial}).
This implies that the plateau is not due to a specific integration method.
In addition, we see in Fig.$\,$\ref{fig:NumAccuracy_errors_64vs80} that the numerical errors decrease by going to the extended precision.

\subsection{Initial state B with the position basis}
\label{app:plateau_x}

\begin{figure}[t]
  \includegraphics[keepaspectratio, scale=0.5]{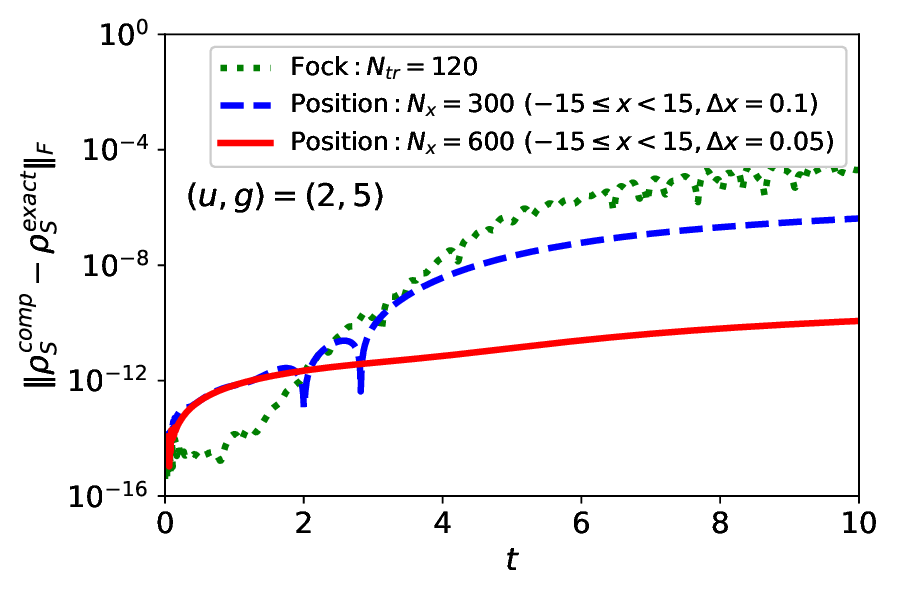}
  \caption{
  Comparison of numerical errors analogous to Fig.$\,$\ref{fig:errors_position_A}, but with the initial two-level state $B$ in Eq.$\,$(\ref{eq:LDiag_Initialtwo-levelAB}).
  }
  \label{fig:errors_position_B}
\end{figure}

In Sec.$\,$\ref{subsec:NumAccuracy_position}, we found that the plateau is not seen when the position basis is used for initial states that satisfy ${\rm tr}_S(\rho_S(0) \sigma_x) = 0$.
When ${\rm tr}_S(\rho_S(0) \sigma_x) \ne 0$, on the other hand, the larger errors remain in the long-time domain.
Figure \ref{fig:errors_position_B} presents a comparison analogous to Fig.$\,$\ref{fig:errors_position_A}, but with the initial state $B$ in Eq.$\,$(\ref{eq:LDiag_Initialtwo-levelAB}).
We see that the error in the long-time domain can be reduced below the plateau value.
By decreasing $\Delta x$, the error can be brought down to the order of $10^{-11} - 10^{-10}$, but not smaller.
These persistent numerical errors in the long-time domain originate from the poor reproduction of ${\rm tr}_S(\rho_S^{\rm comp} \sigma_x)$; we observed that, while ${\rm tr}_S(\rho_S^{\rm comp} \sigma_y) = {\rm tr}_S(\rho_S^{\rm comp} \sigma_z) = 0$ holds in the machine precision order, the value of ${\rm tr}_S(\rho_S^{\rm comp} \sigma_x)$, which is independent of time in the exact solution, constantly increases by about $10^{-14}$ at each time step $0.01$.
We have not yet found a way to resolve this issue.

\section{Numerical sensitivity}
\label{app:cond}

In this appendix, we delve into the origin of the plateau.
Specifically, we reveal the peculiarity of the term involving $u$ in the moment expansion method in Appendix \ref{app:cond_exp}.
In Appendix \ref{app:cond_eig}, we provide a qualitative analysis to address the question of why the plateau is absent with the conventional method.

\subsection{Condition number associated with the matrix exponentiation}
\label{app:cond_exp}

In Appendix \ref{app:plateau_mp}, we find that the plateau can be mitigated by increasing machine precision.
This implies that the errors stem from roundoff errors.
Since the plateau appears regardless of the integration method, the problem of solving the evolution equation $(d/dt) \chi^{\rm rec} = \bar{\mathcal{M}} (\chi^{\rm rec})$ might be itself ill-conditioned, at least in the Fock basis.
A problem is called ill-conditioned when the solution is sensitive to small errors in the input.
In our simulations, we have two inputs that contain roundoff errors; the generator $\bar{\mathcal{M}}$ and the initial condition $\chi^{\rm rec} (t = 0)$.
We here concentrate on the former.

The formal solution to the evolution equation can be obtained by exponentiating the generator $\bar{\mathcal{M}}$.
The exponentiation of $\bar{\mathcal{M}}$ might be sensitive to small errors in it.
This sensitivity can be quantified using a relative condition number associated with the exponentiation.
To understand its meaning, we first consider a simpler quantity; a relative condition number associated with calculating a scalar function $q \to f(q)$.
Assuming $q \ne 0$ and $f(q) \ne 0$, suppose that the input $q$ is perturbed to $q + \delta$.
The relative error in the input then reads $|\delta/q|$ and that in the output reads $|f(q+\delta) - f(q)|/|f(q)| = (|q (d/dq) f(q)|/|f(q)|) \times (\delta / q) + o(\delta) \ (\delta \to 0)$ with $o(\bullet)$ the little-$o$ in the Landau notation.
The ratio of the latter to the former in the limit of small $\delta$, $|q (d/dq) f(q)|/|f(q)|$, quantifies the impact of the input error on the output and is called a relative condition number.
As can be seen from this example, the derivative of a function is needed in evaluating a relative condition number.
For the exponentiation of matrices, thus, the derivative of a matrix function is needed.
For this purpose, one can use the Fr\'echet derivative \cite{Higmfunc}.
An $n$-dimensional matrix function $F$ is said to be Fr\'echet differentiable at a point $Q$ if there exists $L_F(Q,E)$ such that $F(Q+E) - F(Q) - L_F(Q,E) = o(\lVert E \rVert) \ (\lVert E \rVert \to 0)$ for all $n$-dimensional matrices $E$, where $\lVert \bullet \rVert$ is a matrix norm.
The relative condition number can then be introduced as $\lVert Q \rVert \lVert L_F(Q) \rVert / \lVert F(Q) \rVert$ with $\lVert L_F(Q) \rVert = {\rm max}_{E \ne 0} \lVert L_F(Q,E) \rVert / \lVert E \rVert$.
For the matrix exponentiation, we consider $F(Q) = \exp(Q)$.

In practice, we can use the condition number estimator in the SciPy library (${\rm scipy.linalg.expm\_cond}$).
It employs the Frobenius norm as the matrix norm and, for $n$-dimensional matrices, requires construction of an $n^2$-dimensional matrix.
To facilitate the analysis, we need to reduce the dimension of a generator.
To this end,  we note the following identity for any rectangular matrix $\chi^{\rm rec}$ (see Eq.$\,$(\ref{eq:LDian_Mintro}))
\begin{equation*}
  e^{\bar{\mathcal{M}} t} (\chi^{\rm rec}) = \sum_{\alpha,\beta = \pm 1} \ketbra{\alpha}{\beta} \otimes e^{M_{\alpha,\beta}^u t} \braket{\alpha|\chi^{\rm rec}|\beta},
\end{equation*}
where $\ket{\pm1}$ are defined underneath Eq.$\,$(\ref{eq:NumAccuracy_ExactRhoS}) and $M_{\alpha,\beta}^u = - (\kappa/2) a^\dagger a + 2u a^\dagger - ig(\alpha - \beta)(a+a^\dagger)$.
This identity implies that the operators $\{ M_{\alpha,\beta}^u \}_{\alpha,\beta = \pm1}$ can be regarded as generators only on the oscillator system space.
In the long-time domain, only the modes with $\alpha = \beta$ survive (see discussion underneath Eq.$\,$(\ref{eq:LDiag_ExactRhoS})).
Thus, the generator on the oscillator system space responsible for the long-time behavior is given by $M_{\alpha,\alpha}^u = - (\kappa/2) a^\dagger a + 2u a^\dagger$.

\begin{figure}[t]
  \includegraphics[keepaspectratio, scale=0.55]{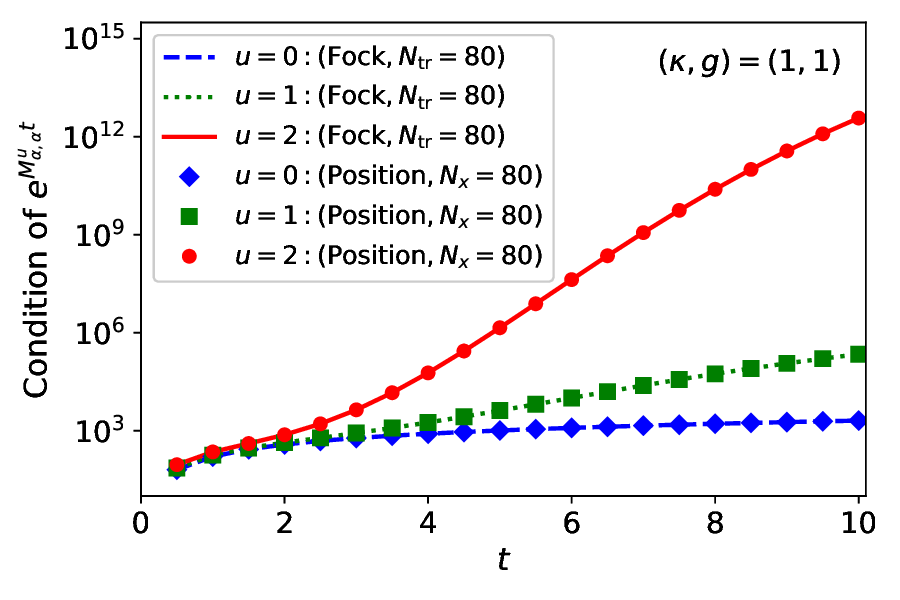}
  \caption{
  Condition number associated with the exponentiation of $M_{\alpha,\alpha}^u t$. We fix $\kappa = 1$ and $g = 1$.
  The lines are the results in the Fock basis and the markers are the results in the position basis.
  In the position basis computation, we set $N_x = 80 \ (-10 \leq x \leq 10, dx = 0.25)$.
  The time $t$ is shown in the unit of $\kappa = 1$.
  }
  \label{fig:NumAccuracy_expmcond_ss}
\end{figure}

The condition numbers associated with the exponentiation of $M_{\alpha,\alpha}^u$ in the Fock basis are shown in Fig.$\,$\ref{fig:NumAccuracy_expmcond_ss}.
When $u = 0$, the generator $M_{\alpha,\alpha}^{u = 0} = - (\kappa/2) a^\dagger a$ is Hermitian.
In this case, the condition number remains almost constant throughout the evolution as can be seen from the dashed blue line.
In contrast, when $u \ne 0$, a marked increase in the condition number is observed.
Thus, the computation of the propagator $\exp(\bar{\mathcal{M}} t)$ with large $t$ might be sensitive to minor errors in $\bar{\mathcal{M}}$ when $u \ne 0$.
Note that a large condition number does not necessarily imply that the associated computation cannot be performed accurately.
At least, Fig.$\,$\ref{fig:NumAccuracy_expmcond_ss} signifies the peculiarity of the term involving $u$ in the generator $\bar{\mathcal{M}}$.

In Sec.$\,$\ref{subsec:NumAccuracy_position}, we saw that numerical accuracy in the long-time domain can be improved by using the position basis. The reason for this is not yet clear.
In Fig.$\,$\ref{fig:NumAccuracy_expmcond_ss}, the same condition number for $M_{\alpha,\alpha}^u$ in the position basis are plotted by the markers.
We see no significant difference between the results in the Fock and position bases.
The difference in numerical accuracy might result from differences in sensitivity to the initial state. This issue might be addressed in future research.

\subsection{Condition number associated with the eigenvalue problem}
\label{app:cond_eig}

Recall that, unlike the moment expansion method, the conventional method does not exhibit a plateau even when $u$ (real part of $\epsilon$) is nonzero.
This difference might be understood from the sensitivity of the matrix exponentiation, according to the discussion in Sec.$\,$\ref{app:cond_exp}.
Since the term involving $u$ in the Liouvillian $\mathcal{L}$ merely generates a unitary transformation, we naively expect it to have no significant effects on the sensitivity.
Quantitatively, we were unable to compute the same condition number for $\mathcal{L}$ within our numerical resource because the dimension is too large.
Instead, we here consider a condition number associated with the eigenvalue problem for the generator.
The relevance of these two condition numbers can be inferred from the theorem 3.15 in \cite{Higmfunc}.
It states that the upper bound of the Fr\'echet derivative, which quantifies the sensitivity of the matrix exponentiation, is proportional to the square of the condition number in the Bauer-Fike theorem, which quantifies the sensitivity of the eigenvalue problem.

A condition number associated with the eigenvalue problem is easier to deal with.
For example, in the moment expansion method, the operator $M_{\alpha,\alpha}^u = - (\kappa/2) a^\dagger a + 2u a^\dagger$ is the relevant generator in the long-time domain when all the eigenvalues of $V_S$ are not degenerate.
This operator is not normal owing to the term involving $u$.
Unlike normal matrices, eigenvalues and eigenvectors of non-normal matrices can be disturbed in their computation by small errors in the input \cite{GVLmcomp}.
It can then be inferred that the term involving $u$ might affect the numerical accuracy in the long-time domain.
This qualitatively agrees with the observations in Sec.$\,$\ref{subsec:NumAccuracy_plateau}.

For more quantitative discussions, here we consider the condition number of a simple eigenvalue introduced in \cite{GVLmcomp},
which is defined as the reciprocal of the inner product of the (normalized) right and left eigenvectors.
We assume that there is no degeneracy in the eigenvalues of $V_S$, as in the two-level target system in Sec.$\,$\ref{sec:NumAccuracy}.
To discuss the long-time behavior, we consider the mode of $M_{\alpha,\alpha}^u$ the eigenvalue of which is zero.
From Eq.$\,$(\ref{eq:LDiag_Mphoton_diag}), the normalized right and left eigenvectors are respectively given by
\begin{equation*}
  \ket{_R \psi_{\rm osc}} = \frac{W_{i,i} \ket{0}}{\sqrt{\braket{0|W_{i,i}^\dagger W_{i,i}|0}}} = e^{- (4u/\kappa)^2/2}e^{(4u/\kappa) a^\dagger} \ket{0},
\end{equation*}
and
\begin{equation*}
  \bra{_L \psi_{\rm osc}} = \frac{ \bra{0} W_{i,i}^{-1} }{\sqrt{\braket{0| W_{i,i}^{-1} (W_{i,i}^{-1})^\dagger |0}}} = \bra{0}.
\end{equation*}
Therefore, the condition number is given by
\begin{equation}
  \braket{_L \psi_{\rm osc}|_R \psi_{\rm osc}}^{-1} =  e^{(4u/\kappa)^2/2}.
  \label{eq:LDiag_Mcond}
\end{equation}
It should be emphasized that this result was obtained without truncating the oscillator state.
Our primary objective is to investigate the sensitivity of numerical simulations where truncation is introduced.
Thus, it is essential to assess the condition number of the truncated generator.
To this end, we numerically computed the right ahd left eigenvectors of the truncated generator $M^u_{\alpha,\alpha}$.
We could then verify the relation Eq.$\,$(\ref{eq:LDiag_Mcond}) for sufficiently large truncation levels.
A deviation from Eq.$\,$(\ref{eq:LDiag_Mcond}) was detected when the right-hand side, $\exp((4u/\kappa)^2/2)$ exceeds $10^{14}$.
This is because the computation of the eigenvectors is ill-conditioned and cannot be performed accurately.

The result Eq.$\,$(\ref{eq:LDiag_Mcond}) implies that the condition number rapidly increases with the value of $u$, which agrees with the behavior in Fig.$\,$\ref{fig:NumAccuracy_expmcond_ss}.
For instance, if $u/\kappa = 2$ as considered in Sec.$\,$\ref{sec:NumAccuracy}, the condition number reads $e^{32} \simeq 10^{14}$.

For comparison, we turn our attention to the identical condition number associated with the Liouvillian $\mathcal{L}$.
We employ the Hilbert-Schmidt inner product to calculate the condition number.
To investigate the behavior in the long-time domain similarly, we examine the mode of $\mathcal{L}_{i,i}$ that possesses an eigenvalue of zero.
We recall that the operation of $\mathcal{L}_{i,i}$ is given by
\begin{equation*}
  \mathcal{L}_{i,i} (\rho) = [(\epsilon - ig_i) a^\dagger - (\epsilon - ig_i)^* a, \rho] + \kappa \mathcal{D}[a] (\rho),
\end{equation*}
which can be diagonalized as Eq.$\,$(\ref{eq:LDiag_W-1LW}) with Eq.$\,$(\ref{eq:LDiag_Wii}).
The right and left eigenoperators are given by
\begin{equation}
  \begin{gathered}
    \mathcal{W}_{i,i} (_R X_{0,0}) \\
    = D \Big( \frac{2(\epsilon - ig_i)}{\kappa} \Big) \ketbra{0}{0} D \Big( \frac{2(\epsilon - ig_i)}{\kappa} \Big)^\dagger,
  \end{gathered}
  \label{eq:LDiag_RightNormalized}
\end{equation}
and
\begin{equation*}
  \Big( \mathcal{W}_{i,i}^{-1} \Big)^{\rm ad} (_L X_{0,0}) = I_{\rm osc},
\end{equation*}
respectively.

These are the results without the truncation.
For the truncated generator, denoted as $\mathcal{L}_{i,i}^{N_{\rm tr}}$ with $N_{\rm tr}$ the truncation level,
we first note that $\mathcal{L}_{i,i}^{N_{\rm tr}}$ still preserves the trace.
This indicates that the $N_{\rm tr}$-dimensional identity operator, denoted by $I_{N_{\rm tr}}$, is a left eigenoperator that has an eigenvalue of zero.
Furthermore, we numerically verified that an eigenvalue of zero is simple and that the corresponding right eigenoperator is given approximately by Eq.$\,$(\ref{eq:LDiag_RightNormalized}) for sufficiently large truncation levels.
In summary, the right eigenoperator of $\mathcal{L}_{i,i}^{N_{\rm tr}}$ with an eigenvalue of zero is a pure state, which we denote as
\begin{equation*}
  _R \rho_{\rm osc}^{N_{\rm tr}} = \ketbra{\varphi_{\rm osc}^{N_{\rm tr}}}{\varphi_{\rm osc}^{N_{\rm tr}}}
\end{equation*}
where $\ket{\varphi_{\rm osc}^{N_{\rm tr}}}$ is a normalized $N_{\rm tr}$-dimensional vector,
whereas the left eigenoperator is the identity operator
\begin{equation*}
  _L \rho_{\rm osc}^{N_{\rm tr}} = \frac{I_{N_{\rm tr}}}{\sqrt{N_{\rm tr}}}.
\end{equation*}
These expressions are normalized as ${\rm tr} ( (_Q \rho_{\rm osc}^{N_{\rm tr}})^\dagger \ _Q \rho_{\rm osc}^{N_{\rm tr}} ) = 1$ for both $Q = R$ and $L$.
From these, the condition number is given by
\begin{equation*}
  [{\rm tr} ( (_L \rho_{\rm osc}^{N_{\rm tr}})^\dagger \ _R \rho_{\rm osc}^{N_{\rm tr}} )]^{-1} = \sqrt{N_{\rm tr}}.
\end{equation*}
Hence, the condition number is independent of the parametrs, in stark contrast to Eq.$\,$(\ref{eq:LDiag_Mcond}).
The formula indicates that the simulation results remain unaffected by the sensitivity within the range of commonly utilized truncation levels $(N_{\rm tr} < 10^5)$.


\end{document}